\documentclass[aps, prd, twocolumn, superscriptaddress, longbibliography]{revtex4-1}

\usepackage{amsmath}
\usepackage{amssymb} 
\usepackage{graphicx}
\usepackage{float}
\usepackage[caption=false]{subfig}
\usepackage{enumerate}
\usepackage{color}
\usepackage{tensor}
\usepackage{mathrsfs}
\usepackage{hyperref}
\usepackage[normalem]{ulem}
\usepackage[dvipsnames]{xcolor}
\usepackage{accents}
\usepackage{scalerel}
\usepackage{bm}
\usepackage{comment}
\DeclareMathOperator{\diag}{diag}

\makeatletter
\newcommand*{\leqdef}{\mathrel{\rlap{%
			\raisebox{0.25ex}{$\m@th\cdot$}}%
		\raisebox{-0.25ex}{$\m@th\cdot$}}%
	=}
\newcommand*{\reqdef}{=\mathrel{\rlap{%
			\raisebox{0.25ex}{$\m@th\cdot$}}%
		\raisebox{-0.25ex}{$\m@th\cdot$}}
}
\makeatother

\begin{document}
	
\title{Influence of gravitational waves upon light in the Minkowski background: from null geodesics to interferometry}
	
\author{Jo\~ao C. Lobato}
\email{jcavlobato@if.ufrj.br}
\affiliation{Universidade Federal do Rio de Janeiro,
		Instituto de F\'\i sica, \\
		CEP 21941-972 Rio de Janeiro, RJ, Brazil}
	
\author{Isabela S. Matos}
\email{isa@if.ufrj.br}
\affiliation{Universidade Federal do Rio de Janeiro,
		Instituto de F\'\i sica, \\
		CEP 21941-972 Rio de Janeiro, RJ, Brazil}
	
\author{Lucas T. Santana}
\email{lts@if.ufrj.br}
\affiliation{Universidade Federal do Rio de Janeiro,
		Instituto de F\'\i sica, \\
		CEP 21941-972 Rio de Janeiro, RJ, Brazil}
	
\author{Ribamar R. R. Reis}
\email{ribamar@if.ufrj.br}
\affiliation{Universidade Federal do Rio de Janeiro,
		Instituto de F\'\i sica, \\
		CEP 21941-972 Rio de Janeiro, RJ, Brazil}
\affiliation{Universidade Federal do Rio de Janeiro, Observat\'orio do Valongo, 
		\\CEP 20080-090 Rio de Janeiro, RJ, Brazil}
	
\author{Maur\'\i cio O. Calv\~ao}
\email{orca@if.ufrj.br}
\affiliation{Universidade Federal do Rio de Janeiro,
		Instituto de F\'\i sica, \\
		CEP 21941-972 Rio de Janeiro, RJ, Brazil}

\begin{abstract}
We have recently derived a manifestly covariant evolution law, under the geometrical optics (or
eikonal) approximation of the vacuum Maxwell's equations, for the electric field along null geodesics
in a general spacetime, relative to an arbitrary set of instantaneous observers \cite{Santana2020}. As one of its
applications, we derive here the final detected intensity signal arising from a prototypical laser interferometric gravitational wave (GW) Michelson-Morley detector, comoving with transverse traceless
(TT) observers, valid for both long and short GW wavelengths (as compared to the lengths of the
interferometer’s arms). The motion of the test particles and light is described through the covariant
kinematic and optical parameters. One of our main results is the presentation of the integrated null
geodesic parametric equations exchanged between two TT observers in terms of explicitly observable
quantities (laser initial frequency and positions of the observers) and the profile of the plane GW
packet. This allows us to revisit the derivation of the consequential radar distance and Doppler shift,
taking the opportunity to discuss some related subtle conceptual issues and how they might affect
the interferometric process. Another achievement is the calculation of the electric field in each arm
up to the detection event, for any relative orientations of the arms and the GW direction. The main
quantitative result is the new expression for the final interference pattern, for normal GW incidence,
which turns out to have three contributions: (i) the well-known traditional one due to the difference
in optical paths, and two new ones due to (ii) the Doppler effect, and (iii) the divergence of the
laser beams. The quantitative relevance of the last two contributions is compared to the traditional
one and shown to be negligible within the geometrical optics regime of light. Although in general
further contributions from the non-parallel transport of the polarization vector are expected (cf.
\cite{Santana2020}), again in the case of GW normal incidence, such a vector is indeed parallel transported, and
those contributions are absent.
\end{abstract}
	
\maketitle
\section{Introduction}
\label{sec:intro}

The first direct detection of gravitational waves (GW) by interferometric experiments \cite{Abbott2016b} heralded a much expected new age of investigation for our Universe. Although the basics regarding interferometry on non-relativistic investigations are well known \cite{Born2002, Hariharan2007}, their relativistic counterparts lead to deep novel conceptual and technical issues \cite{Maggiore2007, Tinto2014, Saulson2017, Bond2016, Reitze2019}. When dealing with this problem, a first attempt is to consider only the interaction of gravity with massive particles, particularly those determining the extremities of the interferometer arms. If this picture is valid, the whole situation can be analyzed as in flat spacetime, with the addition of gravity only as the agent causing the anisotropic stretch of the arms, which furthermore changes light optical paths, providing a non-trivial interference pattern at the end. However, under this perspective, another relativistic aspect which could be relevant is neglected, namely, light's interaction with gravity. 

The metric of spacetime selects the possible null 4-dimensional rays along which light particles travel when the geometrical optics limit of Maxwell's equations is assumed. It is then natural to wonder whether considering such interaction in all its possible facets could bring new elements and corrections to the detection process of GWs. On the one hand, GWs are extremely weak, and we could argue that their effect on light cannot be measured at all. On the other hand, interferometry amplifies small disturbances, and as we shall discuss later on, that interaction is, indeed, already present at linear order in the GW amplitude, the usual control parameter appearing in any modeling of a GW probe. 

Several studies have approached this subject in a variety of aspects, such as: (i) deriving the generic family of null-geodesics in a GW spacetime \cite{Rakhmanov2009, Bini2009, deFelice2010}, (ii) light spatial trajectory perturbations \cite{Finn2009, Kopeikin1999}, (iii) electromagnetic frequency shift \cite{Kaufmann1970, Estabrook1975, Tinto1998, Kopeikin1999, Tinto2002, Armstrong2006} and (iv) electric field evolution along light rays \cite{Santana2020}. Here, we first investigate subjects (i), (ii) and (iii), determining whether the change in luminous spatial path alters the round-trip travel time of light (or equivalently, the radar length of the arm) and what could in principle be the influence of a frequency shift on the phase and intensity of the electromagnetic field. Even though our ultimate concern is with the interferometric procedure, most of the results obtained are valid in a broader picture, where two observers comoving with the transverse traceless (TT) gauge coordinates in a GW spacetime with flat background exchange light rays with each other. We point out that, differently from most of the above mentioned works, here all final quantities are explicitly calculated, not in terms of arbitrary constants of motion, but of known parameters (observables) through the imposition of what we shall call mixed conditions (cf. Eq.~(\ref{mixed_conditions})). 

Those preliminary investigations establish the ground for subject (iv), which is here discussed by the computation of the final electric field and, for normal GW incidence, the detected intensity signal on an idealized interferometric experiment. This usually relies on a Michelson-Morley apparatus, where the intensity pattern is commonly computed \cite{Maggiore2007} in terms of the phase difference between the two beams at the recombination event, being directly related to the difference in radar distances of the two arms. To that end, together with the constancy of phase along light rays, the simple propagation equation, even in TT coordinates, is assumed
\begin{equation}
\frac{dE^{\mu}}{d\vartheta} = 0\,, \label{eq: Mink_Elect_evol}
\end{equation}
where $\vartheta$ is the affine parameter of the null geodesic of choice. In other words, the magnitude and polarization of the electric field $E^{\mu}$ are bound to evolve freely in each arm of the interferometer, as in an inertial frame of Minkowski spacetime, even though GWs are passing by. 

In our recent paper \cite{Santana2020}, as an effort to clear out the laws determining light propagation with respect to an arbitrary set of observers in a generic spacetime, we have shown that the electric field of an electromagnetic (EM) wave in the geometrical optics approximation of Maxwell equations (cf. Appendix \ref{app:geometrical_optics}) evolves along any of its light rays according to
\begin{equation}
\frac{DE^{\mu}}{d\vartheta} + \frac{1}{2} \widehat{\Theta} E^{\mu} = \left( \frac{k^{\mu} E^{\nu} - k^{\nu} E^{\mu}}{\omega_{\textrm{e}}}\right) \frac{Du_{\nu}}{d\vartheta}\,,	
\label{eq:electric_evolution}
\end{equation}
where $\omega_{\textrm{e}}\leqdef - k^{\mu} u_{\mu}$ is the frequency of light, $u^{\mu}$ is the 4-velocity of the frame, $k^{\mu}$ is the null tangent vector of the rays, $\hat{\Theta}$ is the optical expansion of the beam and, of course, for any vector $v^{\mu}$ along the ray, $Dv^{\mu}/d \vartheta \leqdef dv^{\mu}/d \vartheta + \Gamma^{\mu}_{\alpha \gamma} v^{\alpha} k^{\gamma}$. Alternative approaches to describe electric field propagation are~ to choose an auxiliary parallel transported $u^{\mu}$, rendering a vanishing right hand side (RHS) in Eq.~(\ref{eq:electric_evolution}), or to evolve the Faraday tensor instead \cite{Kopeikin2002}. The advantage of the above propagation law is that no posterior boost from an auxiliary observer to the one of interest is needed, and the possible physical effects on the measurable quantity can be assessed even before the solution is obtained.

In contrast to Eq.~(\ref{eq: Mink_Elect_evol}), Eq.~(\ref{eq:electric_evolution}) shows three possibly relevant contributions to the electric field evolution: the connection coefficients on the left-hand side (LHS); the optical expansion and the frame kinematics on the RHS. It becomes expedient, then, to evaluate what is the role played by each of them on the calculation of the interference pattern of a GW detector. The radar distance and frequency shift will appear on the final intensity pattern, giving us the opportunity to bridge our results with subjects (i), (ii) and (iii) and hence provide clarifying insights on the physical origins surrounding the final intensity fluctuations.

In section \ref{sec:models} we describe our model, formalizing our discussion, and comment on an alternative (hybrid) one that is frequently and implicitly adopted in different works, e.g. \cite{Baskaran2004, Rakhmanov2008}, in which perturbations of light spatial trajectories are disregarded. In section \ref{sec:RD_definition}, we evoke the radar distance definition and call special attention to the mixed conditions through which it is uniquely characterized in terms of known quantities. In section \ref{sec:RD_unperturbed_perturbed}, we compute the radar distance within those two models and arrive at the conclusion that they provide the same expression. In section \ref{sec:null_geod}, we obtain the parametric equations of the null geodesics exchanged by TT observers and use kinematic and optical quantities to describe the reference frame and the light beams. In section \ref{sec:Doppler_effect}, we derive the expression for the round-trip frequency shift, interpret it as a Doppler effect and use one of the fundamental laws of electromagnetic geometrical optics to address the common conundrum regarding the possibility of detecting GWs when both arm and light's wavelength are stretched \cite{Faraoni2007, Saulson1997}. In section \ref{sec:interferometry}, the terms appearing in Eq.~(\ref{eq:electric_evolution}) are physically interpreted and the optical expansion and the electric field are propagated along the rays traveling the interferometer arms for any GW incidence and detector configuration. Then, for normal incidence, we obtain the final interference pattern and compare the quantitative relevance of the newly found contributions. In section \ref{sec:conclusion} we discuss our main results and point out further possible developments.

Our signature is $+2$ and we set $c = 1$, unless explicitly stated otherwise. We use Latin letters at the middle of the alphabet to denote generic spatial indices, and at the end of the alphabet to denote specific coordinate indices. Greek component indices can be either spatial or temporal. As concerns the concepts of instantaneous observer, observer, reference frame, photon, we consistently adhere to \cite{Sachs1977}; in particular, a reference frame is conceived as a continuous system and, consequently, its motion can be described similarly to the Newtonian kinematics of an ordinary fluid (see, for example, \cite{Ellis1971}).

\section{Models}
\label{sec:models}
The models we will consider throughout this article to describe light rays propagating in vacuum in a GW spacetime, unless explicitly stated, constitute a two-parameter family defined by the 6-tuple
\begin{equation}
\mathbb{M}_{(\bm{\epsilon})} \leqdef \left(\,{\mathcal N}, \,\varphi, \,g_{(\bm{\epsilon})}, \,u, \,\xi_{(\bm{\epsilon})}, \,\bm{P}\,\right)\,, \label{consistent_models}
\end{equation}
where
\begin{itemize}
	\item $\bm{\epsilon} \leqdef (\epsilon_+,\epsilon_\times)$ is a pair of independent parameters that will be considered up to linear order in all calculations.
	\item ${\mathcal N}$ is a 4-dimensional differentiable manifold;
	\item $\varphi$ is a chart whose coordinate functions $x^{\mu}$ are denoted by $(t, x, y, z)$ and will be called TT coordinates;
	\item $g_{(\bm{\epsilon})}$ is any solution of the linearized vacuum Einstein equations in a Minkowski background, whose components in the $\varphi$ coordinate basis are:
	\begin{align}
	g_{(\bm{\epsilon})\alpha \beta}(t - x) &\leqdef  \eta_{\alpha \beta} + \epsilon_P h^P_{\alpha \beta}(t - x), \label{metric} \\
	\eta_{\alpha\beta} &\leqdef \diag(-1, 1,1,1)\,, \label{Minkowski_matrix}
	\end{align}
	where the two GW polarizations are indexed by $P= +, \times$, for which the usual Einstein summation convention holds, and
	\begin{align}
	\hspace{20pt} h^+_{\alpha \beta}(t-x) & \leqdef h_{+}(t-x)(\delta_{\alpha 3}\delta_{\beta 3} - \delta_{\alpha 2}\delta_{\beta 2})\,, \label{hplus} \\
	\hspace{20pt} h^{\times}_{\alpha \beta}(t-x) & \leqdef -h_{\times}(t-x) (\delta_{\alpha 2} \delta_{\beta 3}+\delta_{\alpha 3} \delta_{\beta 2}) \label{hcross}
	\end{align}
	specify the line element of a GW traveling in the $x$ axis
	\begin{align}
	 \hspace{20pt} ds^2 = & - dt^2 + dx^2 + [1 - \epsilon_+ h_{+}(t - x)]dy^2 \nonumber \\ +& [1 +\epsilon_+ h_{+}(t-x)] dz^2 - 2 \epsilon_{\times} h_{\times}(t-x) dy dz\,. \label{line_elem_TT}
	\end{align}
	
	\item $u$ is the 4-velocity field comoving with $\varphi$, namely,
	\begin{equation}
	u  \leqdef \frac{\partial_t}{\sqrt{-g_{(\bm{\epsilon})}(\partial_t, \partial_t)}} = \partial_t\,.
	\label{TTframe}
	\end{equation}
	This will be called the TT reference frame and, of course, $x^i =$ const along each of its observers.
	
	\item $\xi_{(\bm{\epsilon})}$ is a null geodesic associated to the metric (\ref{metric}), affinely parametrized by $\vartheta$, that is,
	\begin{align}
	g_{(\bm{\epsilon})}(k_{(\bm{\epsilon})}, k_{(\bm{\epsilon})}) & = 0\,, \label{pert_nullity}\\ 
	\frac{D_{(\bm{\epsilon})}}{d\vartheta} k_{(\bm{\epsilon})} & = 0\,, \label{pert_geod}
	\end{align}
	where
	\begin{align}
	k_{(\bm{\epsilon})} \leqdef \frac{d \xi_{(\bm{\epsilon})} }{d\vartheta}\,, \label{pert_velocity}
	\end{align}
	 and $D_{(\bm{\epsilon})}/d\vartheta$ is the directional absolute derivative along the light ray $\xi_{(\bm{\epsilon})}$.
	
	\item $\bm{P}$ is a list of mixed (initial and boundary) data we shall impose on $\xi_{(\bm{\epsilon})}$. Which exact conditions are associated to them and why they are necessary is discussed in section \ref{sec:RD_definition}. 
\end{itemize}

It is immediate to note from the description above that the unperturbed model $\mathbb{M}_{(\bm{0})}$ consists of the Minkowski spacetime as represented through a pseudo-Cartesian coordinate system and its comoving inertial reference frame, whose observers may send light rays to each other via the curves $\xi_{(\bm{0})}$. Every quantity related to this zeroth order model, i.e. those evaluated at $\bm{\epsilon} = \bm{0}$ everywhere, will be  written with a subscript $(\bm{0})$ juxtaposed to its kernel symbol.  In all forthcoming sections, $\bm{\epsilon}$ dependencies in the arguments of functions will be omitted and whenever $\bm{\epsilon}$ is arbitrary so will the sub-indices $(\bm{\epsilon})$.

Moreover, in subsection \ref{subsec:unperturbed_spatial_trajectories} only, a second family of \emph{hybrid} models will be assumed:
\begin{equation}
\mathbb{M}_{\textrm{hyb}(\bm{\epsilon})}\leqdef \left( \mathcal{N},\,\varphi, \,g_{(\bm{\epsilon})}, \,u, \,\tilde{\xi}_{(\bm{\epsilon)}}, \,\bm{P} \right)\,. \label{inconsis_model}
\end{equation} 
The only difference between this family and that provided by Eq.~(\ref{consistent_models}) is simply the curve used to describe light. Here, instead of $\xi_{(\bm{\epsilon})}$, we use the curve $\tilde{\xi}_{(\bm{\epsilon})}$ that satisfies
\begin{equation}
\tilde{\xi}^{i}_{(\bm{\epsilon})} \leqdef \xi_{(\bm{0})}^i \label{xi_tilde}
\end{equation}
and an equation analogous to (\ref{pert_nullity}) for its tangent vectors $\tilde{k}_{(\bm{\epsilon})}$, from which $\tilde{\xi}^t_{(\bm{\epsilon})}$ can be uniquely determined. In other words, $\tilde{\xi}_{(\bm{\epsilon})}$ is a null curve in the current spacetime, with a spatial trajectory coincident with that of the unperturbed curve $\xi_{(\bm{0})}$. As will become explicit later, this second family of models is, apart from very particular circumstances, inconsistent for the description of light propagation since $\tilde{\xi}_{(\bm{\epsilon})}$ is not a geodesic for the metric $g_{(\bm{\epsilon})}$. Nevertheless, Eq.~(\ref{inconsis_model}) will be considered to expose the procedure one would make if light's spatial trajectory perturbations due to its interaction with GWs were neglected, an assumption commonly found in the literature \cite{Baskaran2004, Rakhmanov2008}.  

Last, we emphasize that the metric (\ref{metric}) is a solution of the Einstein's field equations in vacuum, so another possible aspect of the interaction between GWs and electromagnetic waves, namely, the effect of light energy-momentum tensor as a source to the curvature of spacetime will not be considered in this work, that is, light will be held as a test field only (cf. \cite{Schneiter2018}).

\section{Radar distance and mixed conditions} 
\label{sec:RD_definition}

\begin{figure}
	\centering
	\includegraphics[scale=0.15]{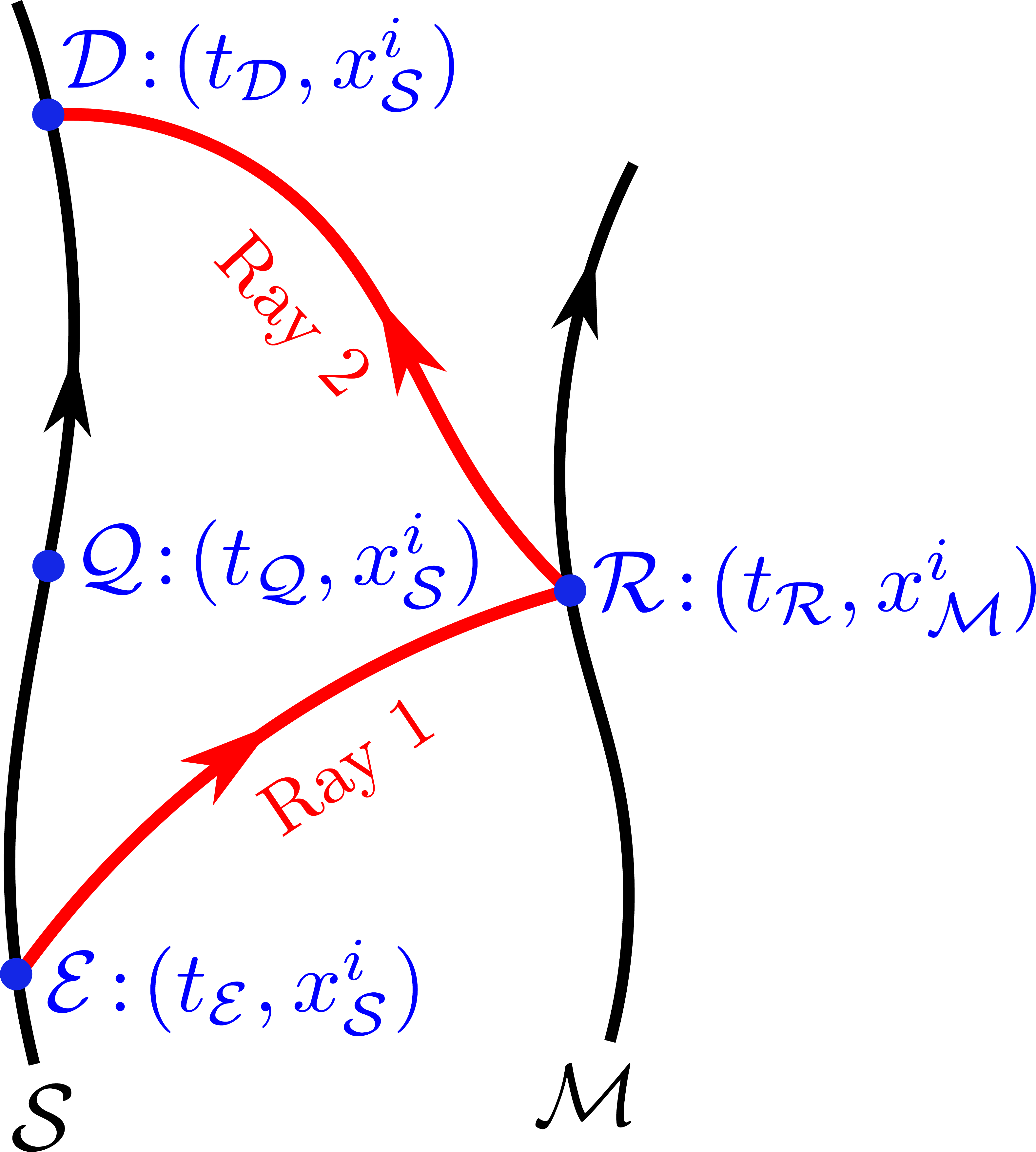} 
	\caption{Observer $\mathcal{S}$ ascribes a radar distance, at its event $\mathcal{Q}$, corresponding to its proper time $t_{\mathcal{Q}}$, to observer $\mathcal{M}$, via null geodesic rays 1 and 2.}
	\label{fig:radar_distance}
\end{figure}

We start by defining the radar distance between two observers of the TT frame, $\mathcal{S}$ (source) and $\mathcal{M}$ (mirror), with the aid of Fig.~\ref{fig:radar_distance}. Observer $\mathcal{S}$ emits a photon from event $\mathcal{E}$ with proper-time $t_{\mathcal{E}}$ (since $g_{tt} = -1$, the proper time of an adapted observer coincides with the coordinate $t$), which travels along ray 1 towards observer $\mathcal{M}$. There, at event $\mathcal{R}$, it is reflected and travels back along ray 2, whereupon it is detected at the event $\mathcal{D}$, with proper time $t_{\mathcal{D}}$.

The radar distance $D_R$ that $\mathcal{S}$ assigns to $\mathcal{M}$ at the ``mid-point'' event $\mathcal{Q}\in\mathcal{S}$, defined to have proper time
\begin{equation}
t_{\mathcal{Q}} \leqdef \frac{t_{\mathcal{E}} + t_{\mathcal{D}}}{2}\,, \label{tQ}
\end{equation}
is
\begin{equation}
D_R(\mathcal{S}, \mathcal{M}, t_{\mathcal{Q}}) \leqdef \frac{t_{\mathcal{D}} - t_{\mathcal{E}}}{2}\,.  \label{D_R_general}
\end{equation}
As defined, such a distance is a purely geometric scalar, with all the needed information being measured by the single observer $\mathcal{S}$. Furthermore it does not rely on a given extended reference frame, as its construction assumes only two observers.

The procedure described above to determine $D_R$ will only be successful once certain boundary conditions are imposed for rays 1 and 2, so that one assures light reaches observer $\mathcal{M}$ and gets back to $\mathcal{S}$. The coordinate representations of rays 1 and 2 will be denoted by $\xi^{\alpha}_{|j}$, where $j = 1,2$ indexes each ray. Then, for each instant $\xi^t_{|1}(0) \reqdef t_{\mathcal{E}}$ of emission (where we chose $\vartheta_{\mathcal{E}}=0$), the partial boundary conditions for ray 1 are:
\begin{align}
&\xi^i_{|1}(0) = x^i_{\mathcal{S}}\,, \label{boundE}\\
&\exists \; \vartheta_{\mathcal{R}} >0 \; | \; \xi^i_{|1}(\vartheta_{\mathcal{R}}) = x^i_{\mathcal{M}}\,, \label{boundR1}
\end{align}
where the fixed spatial coordinates of $\mathcal{S}$ and $\mathcal{M}$ are denoted by $x^i_{\mathcal{S}}$ and $x^i_{\mathcal{M}}$, respectively. These conditions ensure that ray 1 connects events $\mathcal{E}$ and $\mathcal{R}$. For ray 2:
\begin{align}
&\xi^i_{|2}(0) = x^i_{\mathcal{M}}\,, \label{boundR2} \\ 
&\exists \; \vartheta_{\mathcal{D}} >0 \; | \; \xi^i_{|2}(\vartheta_{\mathcal{D}}) = x^i_{\mathcal{S}}\,, \label{boundD}
\end{align}
which guarantees that ray 2 connects events $\mathcal{R}$ and $\mathcal{D}$.

For each emission time $t_{\mathcal{E}}$, these conditions allow one to obtain the radar distance in terms of known quantities. Moreover, they will select, from the family of all possible null geodesics, unique parametrized arcs for rays 1 and 2, provided that an initial value ${\omega_{\textrm{e}}}_{\mathcal{E}}$ for the frequency of light $\omega_{\textrm{e}|1}(\vartheta)$ is given additionally:
\begin{equation}
	{\omega_{\textrm{e}}}_{\mathcal{E}} = \omega_{\textrm{e}|1}(0) = - k_{|1}^{\mu}(0)u_{\mu}(\xi_{|1}(0)) = k_{|1}^t(0)\,. \label{initfreq}
\end{equation}
Of course, the radar distance must be independent of this choice of frequency (achromaticity). Note that, assuming light is emitted from a laser attached to $\mathcal{S}$, since the inner workings of such a device are solely determined by its atomic structure, the initial value of its frequency is not disturbed by the feeble GW, being then independent of $\bm{\epsilon}$. Such a frequency will however evolve throughout the light ray 1 in a non-trivial fashion due to the GW. This will be explored in section \ref{sec:Doppler_effect}. Finally, to connect the frequency of ray 1 at event $\mathcal{R}$ with the initial frequency of ray 2, we assume there occurs a reflection by a mirror at rest on the TT frame, and thus,
\begin{equation}
\omega_{\textrm{e}|1}(\vartheta_{\mathcal{R}}) = \omega_{\textrm{e}|2}(0).\label{continuity}
\end{equation} 

The collection of conditions given by Eqs.~(\ref{boundE} -- \ref{continuity}) will be called \emph{mixed conditions}, and the parameters appearing therein constitute the ingredient
\begin{equation}
	\bm{P} \leqdef \left(\,x^0_{\mathcal E}, x^i_{\mathcal S}, x^i_{\mathcal M}, {\omega_{\textrm{e}}}_{\mathcal{E}}\,\right) \label{mixed_conditions}
\end{equation}
of the models (\ref{consistent_models}). Analogous mixed conditions must be imposed on $\tilde{\xi}$. 

\section{Light spatial trajectories and the radar distance} 
\label{sec:RD_unperturbed_perturbed}

In this section, inspired by the discussions appearing in \cite{Rakhmanov2008, Rakhmanov2009, Finn2009}, we investigate if the spatial trajectory perturbations on the null geodesics due to the GW give rise to linear order corrections on the radar distance between two TT observers. When a GW reaches an interferometer, it changes the radar length of each arm in an anisotropic way, resulting in a non-trivial intensity pattern due to the difference in light travel times in both arms. Should the studied observers $\mathcal{S}$ and $\mathcal{M}$ represent the extremities of the arm of a GW detector, if the radar arm length was calculated neglecting light's spatial trajectory perturbations, such quantity could lead to an incorrect intensity pattern.

\begin{figure}
	\centering
	\includegraphics[scale=0.15]{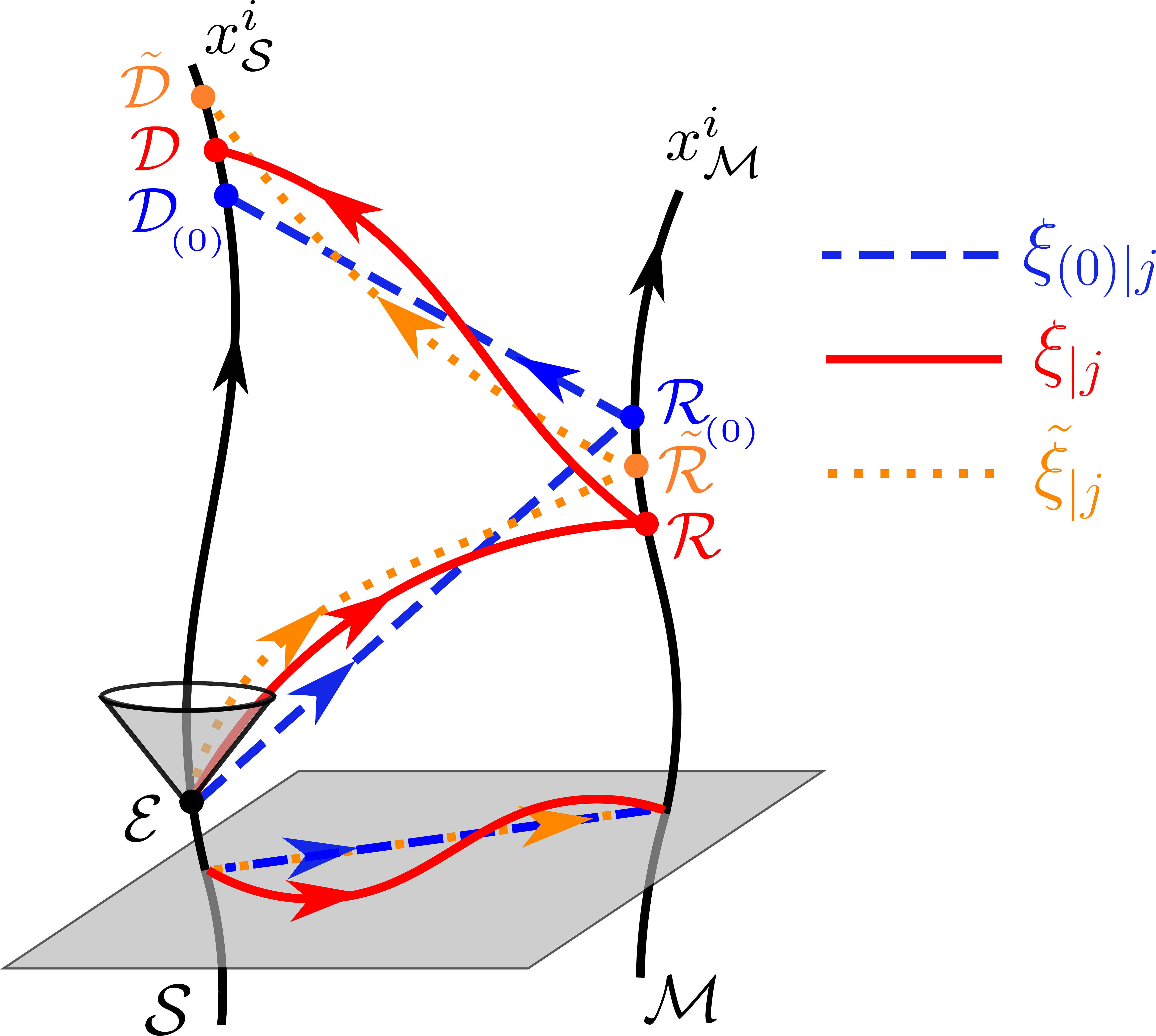}  
	\caption{Minkowski light rays (blue dashed), $\xi_{(\bm{0})|j}$, and perturbed null geodesic rays (red solid), $\xi_{|j}$, connecting two observers. The outgoing arcs ($\xi_{(\bm{0})|1}$ and $\xi_{|1}$, respectively) are constrained to leave observer $\mathcal{S}$ at the same event $\mathcal{E}$ and with the same angular frequency $\omega_{\textrm{e}\mathcal{E}}$; therefore, in general, they reach observer $\mathcal{M}$ at distinct events ($\mathcal{R}_{(\bm{0})}$ and $\mathcal{R}$, respectively) and come back to $\mathcal{S}$, along the incoming arcs ($\xi_{(\bm{0})|2}$ and $\xi_{|2}$, respectively) at distinct events as well ($\mathcal{D}_{(\bm{0})}$ and $\mathcal{D}$, respectively). Besides, we also depict the hybrid null, but not geodesic, curve (orange dotted) $\tilde{\xi}$. Blue and orange curves coincide on the common rest space of $\mathcal{S}$ and $\mathcal{M}$. Whether $\mathcal{R}$ and $\mathcal{\tilde{R}}$ ($\mathcal{D}$ and $\mathcal{\tilde{D}}$) coincide is the subject of section \ref{sec:RD_unperturbed_perturbed}.}
	\label{fig:radar_distance_3_types_of_curve}
\end{figure}

For this discussion, we refer to Fig.~\ref{fig:radar_distance_3_types_of_curve} that shows the configurations of relevant light rays connecting two observers. In order to verify if spatial perturbations actually induce any corrections, we will calculate the radar distance candidate $D_{R, \tilde{\xi}}$ in subsection \ref{subsec:unperturbed_spatial_trajectories}, using the models (\ref{inconsis_model}) and thus the rays $\tilde{\xi}_{|j}$. Then, in subsection \ref{subsec:perturbed_trajectory}\,, the actual radar distance $D_{R, \xi}$ will be calculated within the models (\ref{consistent_models}), and thus using the curves $\xi_{|j}$.  Note that, for both radar distances, we assume the photon to be emitted at the same event $\mathcal{E}$, while events of reflection ($\mathcal{R}$ and $\tilde{\mathcal{R}}$), and reception ($\mathcal{D}$ and $\tilde{\mathcal{D}}$) do not coincide \emph{a priori}. Furthermore, we also begin to compute the null geodesics $\xi$ exchanged by observers $\mathcal{S}$ and $\mathcal{M}$ via integrating its constants of motion, imposing some of the mixed conditions and assuming these constants to depend on the perturbative parameters $\bm{\epsilon}$.

\subsection{Using unperturbed light spatial trajectories} 
\label{subsec:unperturbed_spatial_trajectories}

Here we systematically employ the hybrid models (\ref{inconsis_model}). In this approach, the computation of the radar distance candidate relies on using the curves $\tilde{\xi}$ \cite{Rakhmanov2008, Finn2009}, where
\begin{equation}
\tilde{\xi}^{i}_{|j} (\vartheta) = \xi^{i}_{(\bm{0})|j}(\vartheta) = k^{i}_{(\bm{0})|j} \vartheta + \xi^{i}_{|j}(0)\,. \label{unptraj}
\end{equation}

Since $\tilde{\xi}$ and $\xi_{(\bm{0})}$ obey the partial boundary conditions (\ref{boundE} -- \ref{boundD}), the above equation shows that $\vartheta_{\mathcal{R}_{(\bm{0})}} = \vartheta_{\tilde{\mathcal{R}}}$ and $\vartheta_{\mathcal{D}_{(\bm{0})}} = \vartheta_{\tilde{\mathcal{D}}}$. 

Imposing $ds^2 = 0$ along $\tilde{\xi}_{|1}$ in Eq.~($\ref{line_elem_TT}$), solving for $d\tilde{\xi}^t_{|1}/d\vartheta$ and integrating along ray 1:
\begin{equation}
\Delta t_{\mathcal{E}, \tilde{\mathcal{R}}} = \int^{\vartheta_{\mathcal{R}_{(\bm{0})}}}_{0} \sqrt{[\delta_{ij} + \epsilon_P h^P_{ij}(w_{|1}(\vartheta))] k^i_{(\bm{0})|1} k^j_{(\bm{0})|1}} d\vartheta, \label{timeERunp}
\end{equation}
where, for any pair of events $\mathcal{A}$ and $\mathcal{B}$,
\begin{equation}
\Delta t_{\mathcal{A}, \mathcal{B}} \leqdef t_{\mathcal{B}} - t_{\mathcal{A}}\,,
\end{equation}
and
\begin{align}
w_{|j}(\vartheta) &\leqdef \xi^t_{|j}(\vartheta) - \xi^x_{|j}(\vartheta) \nonumber \\ & = \tilde{\xi}^t_{|j}(\vartheta) - \tilde{\xi}^x_{|j}(\vartheta) + \mathcal{O}(\epsilon_P). \label{w}
\end{align}

Note that $\Delta t_{\mathcal{E}, \tilde{\mathcal{R}}}$ takes into account the perturbations on the metric, though not on the light spatial path, so that it is not the same as $\Delta t_{\mathcal{E}, \mathcal{R}_{(\bm{0})}}$. Moreover, the error in evaluating $\epsilon_P h^P_{ij}$ on $\xi(\vartheta)$ or $\tilde{\xi}(\vartheta)$ is of $\mathcal{O}(\epsilon^2_P)$, since $\tilde{\xi}^t_{(\bm{0})} = \xi^t_{(\bm{0})}$, which justifies the second equality of Eq.~(\ref{w}) as well.  More broadly, we notice that all quantities directly obtained by all the three curves $\xi, \tilde{\xi}$ and $\xi_{(\bm{0})}$ coincide at zeroth order in $\epsilon_P$, a result that will often come in handy in our discussion.

Imposing the previously mentioned partial boundary conditions (\ref{boundE}) and (\ref{boundR1}) to $\xi_{(\bm{0})|1}$ on Eq.~(\ref{unptraj}) and remembering that $\eta_{\mu \nu} k^{\mu}_{(\bm{0})}k^{\nu}_{(\bm{0})} = 0$: 
\begin{equation}
k^{i}_{(\bm{0})|1} = \frac{\Delta x^i}{\vartheta_{\mathcal{R}_{(\bm{0})}}}\,, \quad k^t_{(\bm{0})|1} = \frac{\Delta \ell}{\vartheta_{\mathcal{R}_{(\bm{0})}}}\,, \label{k_{(0)}}
\end{equation}
where
\begin{align}
\Delta x^i \leqdef x^i_{\mathcal{M}} - x^i_{\mathcal{S}}\,, \quad \Delta \ell \leqdef \sqrt{\delta_{ij}\Delta x^i \Delta x^j}\,. \label{delta_l}
\end{align}
Expanding (\ref{timeERunp}) and using (\ref{k_{(0)}}):
\begin{equation}
\Delta t_{\mathcal{E}, \tilde{\mathcal{R}}} = \Delta \ell + \frac{\Delta x^i \Delta x^j}{2 \vartheta_{\mathcal{R}} \Delta \ell} \int_{0}^{\vartheta_{\mathcal{R}}} \epsilon_P h^P_{ij} (w_{|1}(\vartheta))  d\vartheta,
\end{equation}
where $\vartheta_{\mathcal{R}_{(\bm{0})}}$ was replaced by $\vartheta_{\mathcal{R}}$ in the second term with an error of $\mathcal{O}(\epsilon^2_P)$. Now, we change the integration variable to 
\begin{align}
w_{(\bm{0})|1}(\vartheta) &=  (k^t_{(\bm{0})|1}-k^x_{(\bm{0})|1}) \vartheta + t_{\mathcal{E}} - x_{\mathcal{S}}, \label{u_(0)}
\end{align}
and, then, defining
\begin{equation}
m_{P|j}(\vartheta) \leqdef \int_{w_{|j}(0)}^{w_{|j}(\vartheta)} h_{P}(w)\, dw , \label{n1}
\end{equation}
and remembering Eqs.~(\ref{hplus}) and (\ref{hcross}), we arrive at
\begin{align}
\Delta t_{\mathcal{E}, \tilde{\mathcal{R}}} = \Delta \ell - \frac{1}{2(\Delta \ell - \Delta x)\Delta \ell}\big[ 2\epsilon_{\times} \Delta y \Delta z\,   m_{\times|1}(\vartheta_{\mathcal{R}})& \nonumber \\ + \epsilon_+(\Delta y^2 - \Delta z^2)\,  m_{+|1}(\vartheta_{\mathcal{R}})  &\big]. \label{tR-tE}
\end{align}
A similar computation for the back-trip gives:
\begin{align}
\Delta t_{\tilde{\mathcal{R}}, \tilde{\mathcal{D}}} =  \Delta \ell - \frac{1}{{2(\Delta \ell + \Delta x)\Delta \ell}} \big[ 2\epsilon_{\times} \Delta y \Delta z \,  m_{\times|2}(\vartheta_{\mathcal{D}})& \nonumber \\ + \epsilon_+ (\Delta y^2 - \Delta z^2)\,  m_{+|2}(\vartheta_{\mathcal{D}}) &\big]\,. \label{tD-tR}
\end{align}

Finally, adding Eqs.~(\ref{tR-tE}) and (\ref{tD-tR}), the radar distance candidate between $\mathcal{S}$ and $\mathcal{M}$ at the mid-point event $\tilde{Q}$, $D_{R,\tilde{\xi}} \leqdef \Delta t_{\mathcal{E}, \mathcal{\tilde{D}}}/2$, is found. A useful perspective, for example, in identifying the radar Doppler effect (cf. section \ref{sec:Doppler_effect}), is to admit, moreover, that $\mathcal{S}$ emits (and also detects) photons at all times, probing the distance $D_{R,\tilde{\xi}} (t_{\mathcal{Q}})$ for all instants $t_{\mathcal{Q}}$ in its worldline. To make explicit its dependence with the time coordinate $t_{\mathcal{Q}}$, it is only necessary to note that,
\begin{align}
\epsilon_P m_{P|1}(\vartheta_{\mathcal{R}}) &= \epsilon_P \int^{t_{\mathcal{R}}- x_{\mathcal{M}}}_{t_{\mathcal{E}} - x_{\mathcal{S}}} h_{P}(w) dw \nonumber \\ &= \epsilon_P \int^{t_{\mathcal{Q}}- x_{\mathcal{M}}}_{t_{\mathcal{Q}} - \Delta\ell - x_{\mathcal{S}}} h_{P}(w) dw \reqdef \epsilon_P M_{P|1} (t_{\mathcal{Q}})\,, \label{M_P_1}
\end{align}
and similarly:
\begin{align}
\epsilon_P m_{P|2}(\vartheta_{\mathcal{D}}) &=  \epsilon_P \int^{t_{\mathcal{D}}- x_{\mathcal{S}}}_{t_{\mathcal{Q}} - x_{\mathcal{M}}} h_{P}(w) dw \nonumber \\ &= \epsilon_P \int^{t_{\mathcal{Q}}+\Delta \ell -x_{\mathcal{S}}}_{t_{\mathcal{Q}}  - x_{\mathcal{M}}} h_{P}(w) dw \reqdef \epsilon_P M_{P|2} (t_{\mathcal{Q}})\,. \label{M_P_2}
\end{align}
Then 
\begin{align}
&D_{R,\tilde{\xi}} (t_{\mathcal{Q}}) = \Delta \ell \,- \nonumber \\ & \frac{1}{2\Delta \ell} \left\{\epsilon_+ \frac{\Delta y^2 - \Delta z^2}{2} \left[\frac{ M_{+|1} (t_{\mathcal{Q}})}{\Delta \ell - \Delta x} +\frac{ M_{+|2}(t_{\mathcal{Q}})}{\Delta \ell + \Delta x} \right]  \right. + \nonumber \\
& \hspace{50pt} \left.  \epsilon_{\times} \Delta y \Delta z \left[\frac{ M_{\times|1}(t_{\mathcal{Q}})}{\Delta \ell - \Delta x} + \frac{ M_{\times|2}(t_{\mathcal{Q}})}{\Delta \ell + \Delta x} \right] \right\}. \label{D_R}
\end{align}

As pointed out by \cite{Finn2009}, the procedure adopted in this subsection, although commonly used when discussing an interferometer response to GWs \cite{Baskaran2004, Rakhmanov2008}, seems inconsistent. What, then, is the interpretation of (\ref{D_R})? It gives the radar distance if a photon could still travel along its unperturbed spatial trajectory, obeying Eq.~(\ref{unptraj}), even with the presence of GWs, but with the zeroth component of its parametric curve perturbed by them. However, that proves not to be possible, since $\tilde{\xi}_{|j}$, even though satisfying $ds^2=0$, cannot be additionally a geodesic, in general, when (\ref{line_elem_TT}) is assumed. Indeed, from the solutions for $\xi_{|1}$ that will be obtained in subsection \ref{subsec:perturbed_trajectory}\,,  $\xi^{i}_{|j} \neq \tilde{\xi}^i_{|j}$. Therefore, it is imperative to compute the fully linearly perturbed null geodesics from observer $\mathcal{S}$ to observer $\mathcal{M}$ (i.e. obeying the necessary partial boundary conditions), to rigorously derive a physically meaningful radar distance. 

The remaining question to be answered is \cite{Rakhmanov2009, Finn2009}: can those trajectory perturbations significantly alter the distance traveled by light? Certainly, the simplifying assumption made in this section forbids a proper analysis of the electromagnetic frequency evolution along the light ray taking the GW into account as we will show in section \ref{sec:Doppler_effect} (cf. also \cite{Kaufmann1970}). Using Eqs.~(\ref{pert_nullity}) and (\ref{k_{(0)}}) together with the expansion
\begin{equation}
	k^i(\vartheta) \reqdef k^i_{(\bm{0})}(\vartheta) + \epsilon_P k^{iP}(\vartheta), \label{k_expansion}
\end{equation}
it is easily seen that, for the TT frame,
\begin{equation}
	\omega_{\textrm{e}}(\vartheta) = k^t(\vartheta) = \tilde{\omega}_{\textrm{e}}(\vartheta) +  \frac{\epsilon_P}{2 \Delta \ell}\delta_{ij} \Delta x^j k^{iP}(\vartheta)\,,
\end{equation}
where $\tilde{\omega}_\textrm{e} \leqdef -u_\mu \tilde{k}^\mu$. We can then expect conceptual and practical aspects regarding this interaction that cannot be attainable through the hybrid model from Eq.~(\ref{inconsis_model}).
\subsection{Using perturbed light spatial trajectory}
\label{subsec:perturbed_trajectory}
To obtain the actual null geodesics of this GW spacetime, it is useful to change coordinates to
\begin{align}
u \leqdef \frac{t - x}{2}, \quad
v \leqdef \frac{t + x}{2}, \quad y \leqdef y, \quad z \leqdef z. \label{null_coordinates}
\end{align}
In this new coordinate system, the metric components do not depend on $v$, $y$ and $z$, so that there exist three constants of motion along the light rays:
\begin{align}
&\delta \leqdef k_v, \label{kv} \\
&\alpha \leqdef k_y, \label{ky}\\
&\beta\leqdef k_z. \label{kz}
\end{align} 
From these three constants, one may use the linearized inverse metric to obtain
\begin{align}
	&k^u = - \frac{\delta}{2} \label{k^u}\,, \\  
	&k^y = (1+h_+) \alpha + h_{\times} \beta \label{k^y}\,, \\ 
	&k^z = (1-h_+) \beta + h_{\times} \alpha\,. \label{k^z}
\end{align}
Using $k^{\mu}k_{\mu} = 0$: 
\begin{equation}
	2 k^v \delta = - [(1+h_+)\alpha^2 + (1-h_+)\beta^2 + 2 \alpha \beta h_{\times}]. \label{k^v}
\end{equation}
Integrating $k^{u}$:
\begin{equation}
w(\vartheta) \leqdef \xi^t(\vartheta) - \xi^x(\vartheta) =  - \vartheta \delta + t_{\mathcal{E}} - x_{\mathcal{S}}. \label{u}
\end{equation}
Then, for $\delta \neq 0$ \cite{Rakhmanov2009, deFelice2010, Bini2009}, we find the remaining component $k^v$, and the parametric equations associated to the light rays in terms of the constants of motion are determined by integrating Eqs.~(\ref{k^y})--(\ref{k^v}), together with Eq.~(\ref{u}). They can be expressed in terms of $m_P$, defined in Eq.~(\ref{n1}), as:
\begin{align}
\xi^t(\vartheta) = \,& \xi^t(0) - \frac{1}{2} \vartheta\,  (1 + A^2 + B^2)\,\delta + \nonumber \\ & \frac{1}{2} \,\epsilon_+\, m_+ (\vartheta) (A^2 - B^2) + \epsilon_{\times}\, m_{\times}(\vartheta) AB\,, \label{t} \\
\xi^x(\vartheta) = \,& \xi^x(0) + \frac{1}{2} \vartheta\,(1 - A^2 - B^2)\,\delta + \nonumber \\ &  \frac{1}{2} \, \epsilon_+\, m_+ (\vartheta)  ( A^2 - B^2) + \, \epsilon_{\times}\, m_{\times}(\vartheta) AB\,, \label{x} \\[5pt]
\xi^y(\vartheta) = \,& \xi^y(0) + \vartheta A\,\delta  - A\, \epsilon_+\, m_+(\vartheta) - B\, \epsilon_{\times}\, m_{\times}(\vartheta) \,, \label{y} \\[5pt]
\xi^z(\vartheta) = \,& \xi^z(0) + \vartheta B\,\delta  + B\, \epsilon_+\, m_+(\vartheta) - A\, \epsilon_{\times}\, m_{\times}(\vartheta)\,, \label{z}
\end{align}
where 
\begin{align}
	A\leqdef \alpha/\delta\,, \quad
	B\leqdef \beta/\delta\,.
	\label{eq:constants_of_motion_A_B}
\end{align}

For $\delta = 0$, Eq.~(\ref{k^v}) leads to a solution for $\alpha$ in terms of $\beta$:
\begin{equation}
\alpha = \frac{-\beta \epsilon_{\times}\, h_{\times} \pm i |\beta|}{1 + \epsilon_+\, h_+}\,,
\end{equation}
for which the only possible real solution is $\alpha= \beta = 0$. Replacing $\delta=\alpha=\beta=0$ in Eqs.~(\ref{k^y}), (\ref{k^z}) and (\ref{u}) and integrating:
\begin{align}
&\xi^y(\vartheta) = \xi^y(0)\,, \quad \xi^z(\vartheta) = \xi^z(0)\,, \\
& \hspace{15pt} \xi^x(\vartheta)  = \xi^t(\vartheta) - t_{\mathcal{E}} + x_{\mathcal{S}}\,. \label{xdelta0}
\end{align}
Finally, noting that $\Gamma^t_{\mu \nu} = 0$ for $\mu,\nu = t, x$ and using the explicit geodesic equation for $\xi^t$:
\begin{equation}
\xi^t(\vartheta) = C \vartheta + t_{\mathcal{E}}\,,
\end{equation}
where $C$ is a constant. Once substituted in Eq.~(\ref{xdelta0}):
\begin{equation}
\xi^x(\vartheta) = C\vartheta + x_{\mathcal{S}}\,.
\end{equation} 

We note that, in \cite{Rakhmanov2014}, the $\delta = 0$ solution is attributed only to non-null geodesics, while we have shown its existence for null geodesics as well. It is the trajectory of a photon traveling purely along the GW propagation direction (here, the $x$ direction). We notice that this trajectory is not affected by the GW in any way. For this special case, the procedure used to calculate the radar distance in subsection \ref{subsec:unperturbed_spatial_trajectories} is in fact rigorous. A solution with this property is expected, for example, by Lemma II found in p. 326 of \cite{Rindler2006}. 

Since there is no perturbation in the $\delta = 0$ case, the focus here will be on the $\delta \neq 0$ solutions. If our sole intention is to calculate the complete radar distance, there is no need to solve for $A$, $B$ and $\delta$ when imposing the mixed conditions to the null geodesics, as exemplified by the treatment in \cite{Rakhmanov2009}. Here, however, the value of these constants will be explicitly computed so that the general parametric equations for rays 1 and 2 can be obtained and further aspects of the influence of GWs on light (for example, how the frequency evolves along the rays, in section \ref{sec:Doppler_effect}\,) can be thoroughly analyzed. 

The system arising from evaluating Eqs.~(\ref{x})--(\ref{z}) at $\vartheta_{\mathcal{R}}$, imposing the partial boundary conditions for ray 1, Eqs.~(\ref{boundE}) and (\ref{boundR1}), and using Eq.~(\ref{u}) to replace $ \vartheta_{\mathcal{R}} \delta$ for $\Delta x - \Delta t_{\mathcal{E, R}}$ results in:
\begin{align}
\Delta x &= \frac{1}{2} [(\Delta x - \Delta t_{\mathcal{E}, \mathcal{R}}) (1 - A^2_{|1} - B^2_{|1}) + \nonumber \\ &  \hspace{23pt} 2\, \epsilon_{\times}\, m_{\times|1}(\vartheta_{\mathcal{R}}) A_{|1} B_{|1}+ \nonumber \\ & \hspace{30pt} \epsilon_+\, m_{+|1}(\vartheta_{\mathcal{R}})  (A^2_{|1} - B^2_{|1}) ]\,, \label{xbound} \\
\Delta y &= A_{|1} (\Delta x - \Delta t_{\mathcal{E}, \mathcal{R}}) - A_{|1} \epsilon_+\, m_{+|1}(\vartheta_{\mathcal{R}}) - \nonumber \\ & \hspace{13pt} B_{|1} \epsilon_{\times}\, m_{\times|1}(\vartheta_{\mathcal{R}})\,,  \label{ybound} \\
\Delta z &=  B_{|1} (\Delta x - \Delta t_{\mathcal{E}, \mathcal{R}}) + B_{|1} \epsilon_+\, m_{+|1}(\vartheta_{\mathcal{R}}) \,- \nonumber \\ & \hspace{13pt}A_{|1} \epsilon_{\times}\, m_{\times|1}(\vartheta_{\mathcal{R}})\,. \label{zbound}
\end{align}
Combining Eqs.~(\ref{ybound}) and (\ref{zbound}) yields:
\begin{align}
&\epsilon_+ m_{+|1}(\vartheta_{\mathcal{R}}) \frac{(A^2_{|1} - B^2_{|1})}{2} = -\frac{(A_{|1} \Delta y + B_{|1} \Delta z)}{2} +\nonumber \\ & (\Delta x - \Delta t_{\mathcal{E}, \mathcal{R}})\frac{(A^2_{|1} + B^2_{|1})}{2} - \epsilon_{\times} m_{\times |1}(\vartheta_{\mathcal{R}})A_{|1} B_{|1} \,. 
\end{align}
Substituting for this into Eq.~(\ref{xbound}) provides
\begin{equation}
\Delta t_{\mathcal{E, R}} = - (\Delta x + A_{|1} \Delta y + B_{|1} \Delta z)\,. \label{timeAB}
\end{equation}
Replacing this last expression in Eqs.~(\ref{ybound}) and (\ref{zbound}), a system of two equations and two unknowns ($A_{|1}$ and $B_{|1}$) is attained. It will be solved perturbatively by writing 
\begin{align}
	A_{|1} = A_{(\bm{0})|1} + A_{+|1} \epsilon_+ + A_{\times|1} \epsilon_{\times}\,,
	\label{eq:A_perturbative_expansion}
\end{align}
and a similar expression for $B_{|1}$.  

In zeroth order ($\epsilon_+ = \epsilon_{\times} = 0$):
\begin{align}
&2A_{(\bm{0})|1} \Delta x + (A_{(\bm{0})|1}^2 - 1) \Delta y + A_{(\bm{0})|1} B_{(\bm{0})|1} \Delta z = 0\,, \label{A_0^2} \\
&2B_{(\bm{0})|1} \Delta x + A_{(\bm{0})|1} B_{(\bm{0})|1} \Delta y + (B_{(\bm{0})|1}^2 - 1) \Delta z = 0\,, \label{B_0^2}
\end{align}
whose solutions are:
\begin{equation}
A_{(\bm{0})|1} = \frac{\Delta y}{\Delta x - \Delta \ell} \;\; \text{and} \;\; B_{(\bm{0})|1} = \frac{\Delta z}{\Delta x - \Delta \ell}\,. 
\end{equation}

In first order, since $\epsilon_+$ and $\epsilon_{\times}$ are considered independent, we can solve for their corresponding terms separately. For plus polarization ($\epsilon_{\times} = 0$), replacing the values for $A_{(\bm{0})|1}$ and $B_{(\bm{0})|1}$, the system becomes 
\begin{equation}
	G \begin{bmatrix}
	A_{+|1} \\ B_{+|1}
	\end{bmatrix} = \begin{bmatrix}
	\Delta y \\ - \Delta z
	\end{bmatrix} m_{+|1} (\vartheta_{\mathcal{R}}),
\end{equation}
where
\begin{equation}
	G \leqdef \begin{bmatrix}
	(\Delta x - \Delta \ell)^2 + \Delta y^2 & \Delta y \Delta z \\ \Delta y \Delta z & (\Delta x - \Delta \ell)^2 + \Delta z^2
	\end{bmatrix}.
\end{equation}
Analogously, with only the cross polarization ($\epsilon_+=0$):
\begin{equation}
G \begin{bmatrix}
A_{\times|1} \\ B_{\times|1}
\end{bmatrix} = \begin{bmatrix}
\Delta z \\ \Delta y
\end{bmatrix} m_{\times|1} (\vartheta_{\mathcal{R}}).
\end{equation} 
Inverting these linear systems, the complete solutions for $A_{|1}$ and $B_{|1}$ are found to be:
\begin{align}
A_{|1} =\ & \frac{1}{\Delta x - \Delta \ell}\bigg\{\Delta y - \nonumber \\ & \hspace{10pt} \frac{1}{2\Delta \ell} \bigg[\Delta y \bigg(1 + \frac{2\Delta z^2}{(\Delta x - \Delta \ell)^2} \bigg) \epsilon_+\, m_{+|1}(\vartheta_{\mathcal{R}}) + \nonumber \\ 
&\hspace{35pt}  \Delta z \bigg(1 + \frac{\Delta z^2 - \Delta y^2}{(\Delta x - \Delta \ell)^2} \bigg) \epsilon_{\times}\, m_{\times|1}(\vartheta_{\mathcal{R}}) \bigg]\bigg \}\,, \label{A1}\\
B_{|1}=\ & \frac{1}{\Delta x - \Delta \ell}\bigg\{\Delta z + \nonumber \\ &  \hspace{10pt}\frac{1}{2\Delta \ell} \bigg[\Delta z \bigg(1 + \frac{2\Delta y^2}{(\Delta x - \Delta \ell)^2} \bigg) \epsilon_+\, m_{+|1}(\vartheta_{\mathcal{R}}) - \nonumber \\ 
&\hspace{35pt} \Delta y \bigg(1 + \frac{\Delta y^2 - \Delta z^2}{(\Delta x - \Delta \ell)^2} \bigg) \epsilon_{\times}\, m_{\times|1}(\vartheta_{\mathcal{R}}) \bigg]\bigg \}\,. \label{B1}
\end{align}
Replacing them on Eq.~(\ref{timeAB}) and comparing to Eq.~(\ref{tR-tE}):
\begin{equation}
	\Delta t_{\mathcal{E, R}} = \Delta t_{\mathcal{E}, \tilde{\mathcal{R}}}. \label{sametime}
\end{equation}
By making the changes $\Delta x^i \rightarrow -\Delta x^i$ and   $m_{P|1}(\vartheta_{\mathcal{R}}) \rightarrow m_{P|2}(\vartheta_{\mathcal{D}})$, we conclude that the back-trip times also coincide. As a consequence: 
\begin{equation}
	D_{R,\xi} = D_{R,\tilde{\xi}}. \label{eq:coincidence}
\end{equation}

This may sound as an unexpected coincidence, since the spatial trajectories of light in Minkowski spacetime and those in a GW one are not, in general, the same. In \cite{Rakhmanov2009}, this coincidence is noticed but no discussion is made on which hypotheses are behind such a result. In \cite{Finn2009}, a set of mathematical conditions for the validity of Eq.~(\ref{eq:coincidence}) is presented in a more general context justifying such result in our case. We conclude that, indeed, the perturbations in light spatial trajectories do not alter the linearized travel time, although being relevant to other luminous related phenomena, e.g. the changes in electromagnetic frequency along light rays and the electric field itself, as will be discussed further.

The radar distance expression computed in this section will be important when we consider the evolution of the electric field within a toy GW detector. We can rewrite it by introducing angles $\theta$ and $\phi$ such that: 
\begin{align}
&\Delta x = \Delta \ell \cos{\theta}\,, \quad\Delta y = \Delta \ell \sin{\theta} \cos{\phi}\,, \quad \nonumber \\ &\hspace{35pt}\Delta z = \Delta \ell \sin{\theta} \sin{\phi}\,. \label{sph} 
\end{align}

\begin{figure}[t]
	\centering
	\includegraphics[scale = 0.15]{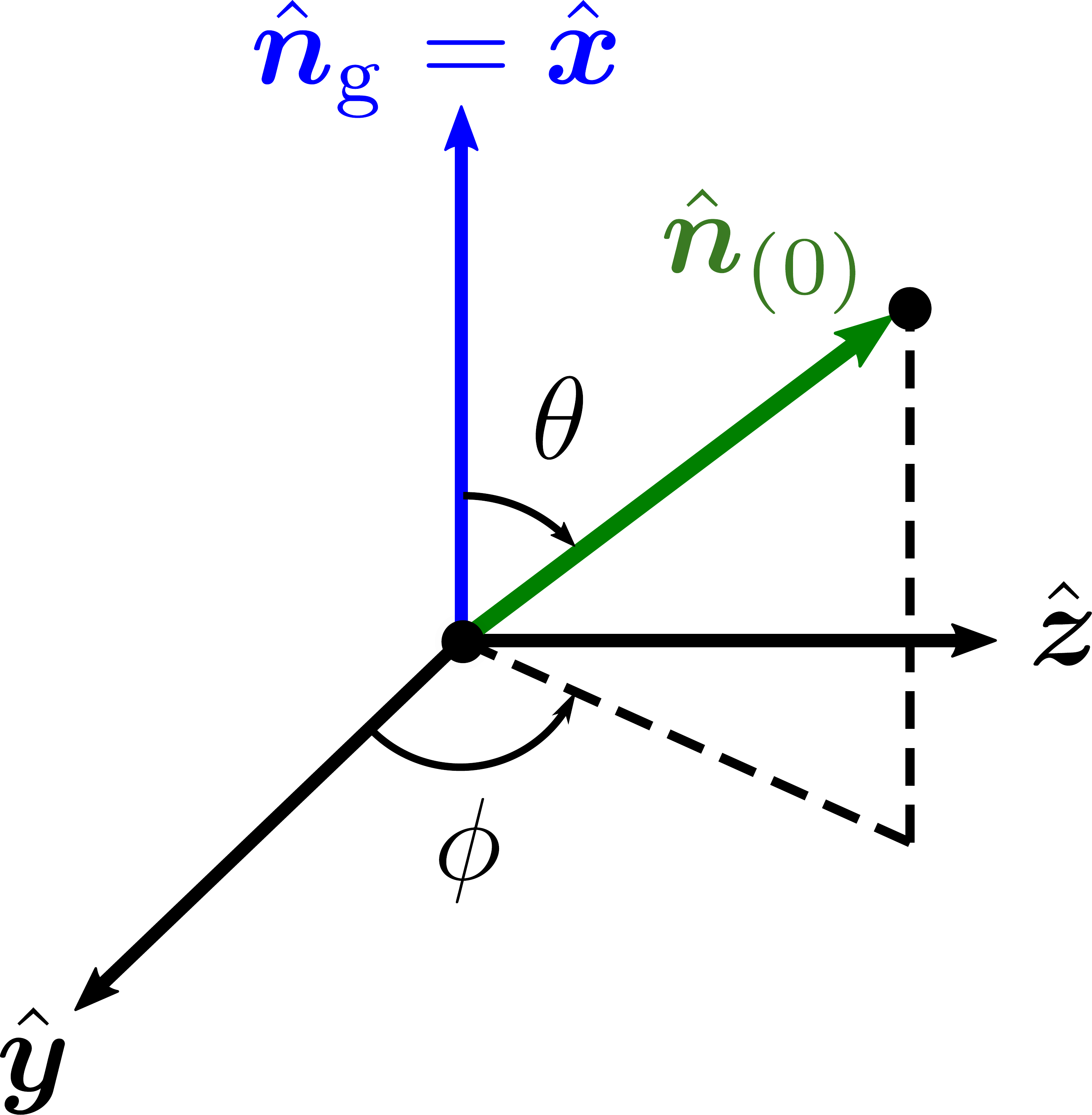}
	\caption{Three-dimensional rest space of an instantaneous observer immersed in the GW region. The spatial direction of propagation of the gravitational wave, {$\hat{\boldsymbol n}_{\textrm{g}}$}, is chosen along the local unit vector $\hat{{\boldsymbol x}}$, whereas the local zeroth order spatial direction of the laser beam is $\hat{\boldsymbol n}_{(\bm{0})}$\,.} 
	\label{fig:arm_direction}
\end{figure}
Replacing this in Eq.~(\ref{D_R}) gives:
\begin{align}
&D_{R, \xi}(t_{\mathcal{Q}}) =\, \Delta \ell- 
 \nonumber \\ &\frac{\epsilon_+}{2} \left[ \cos^2\left(\frac{\theta}{2}\right)M_{+|1}(t_{\mathcal Q}) +  \sin^2\left( \frac{\theta}{2}\right) M_{+|2}(t_{\mathcal Q}) \right]\cos 2\phi {} -\nonumber \\
&  \frac{\epsilon_\times}{2} \left[ \cos^2\left( \frac{\theta}{2} \right)M_{\times|1}(t_{\mathcal Q}) + \sin^2\left( \frac{\theta}{2} \right)M_{\times|2}(t_{\mathcal Q}) \right]\sin 2\phi \,. \label{radardistspheric}
\end{align}
If observers $\mathcal{S}$ and $\mathcal{M}$ are the extremities of an interferometer arm, $\theta$ is the angle between the GW propagation and the zeroth order arm directions, as depicted in Fig.~\ref{fig:arm_direction}.
  
\section{Null geodesics and TT frame kinematics} \label{sec:null_geod}

From now on, only the perturbed, red, solid curves of Fig.~\ref{fig:radar_distance_3_types_of_curve} will be used. In this section, we begin by determining the remaining constants of motion and the general parametric expressions for the null rays (subsection \ref{subsec: gen_parametric_eq}). This result will enable us to determine the evolution of the frequency along light rays on section (\ref{sec:Doppler_effect}) and will be useful latter when solving the electric field evolution.

Although the previous discussions assumed only two null geodesic arcs being exchanged by $\mathcal{S}$ and $\mathcal{M}$, we will always have in mind that these rays are part of a null congruence of curves (a beam of light). We address, then, on subsection \ref{subsec:kin_parameters}, two elements present in Eq.~(\ref{eq:electric_evolution}), namely, the kinematics of the TT reference frame and the optical parameters characterizing the light beam.

\subsection{The general parametric equations for ray 1}  \label{subsec: gen_parametric_eq}

The initial frequency measured by $\mathcal{S}$ is given by Eq.~(\ref{initfreq}), which, differentiating Eq.~(\ref{t}) with respect to $\vartheta$ and evaluating in $\mathcal{E}$ becomes:
\begin{align}
&{\omega_{\textrm{e}}}_{\mathcal{E}} = - \frac{\delta_{|1}}{2} [1 + A^2_{|1} + B^2_{|1} + 2\epsilon_{\times}{h_{\times|1}}(0)A_{|1}B_{|1} +\nonumber \\ & \hspace{57pt} \epsilon_+(A^2_{|1} - B^2_{|1}) h_{+|1}(0)],  \label{init_freq}
\end{align}
where $h_{P|j}(\vartheta) \leqdef h_P(\xi^t_{|j}(\vartheta) - \xi^x_{|j}(\vartheta)) $, and Eq.~(\ref{u}) was used to calculate the derivatives of $m_{P|j}$:
\begin{equation}
\frac{dm_{P|j}}{d \vartheta} = - h_{P|j}  \delta_{|j}. \label{np_deriv}
\end{equation}
Expression (\ref{init_freq}) can be inverted to give the value of the constant $\delta_{|1}$. Replacing Eqs.~(\ref{A1}) and (\ref{B1}) subsequently:
\begin{align}
&\delta_{|1} = - {\omega_{\textrm{e}}}_{\mathcal{E}} \bigg\{1 - \frac{\Delta x}{\Delta\ell}+ \nonumber \\ & \frac{1}{\Delta\ell^3} \bigg[\epsilon_+ \left(m_{+|1}(\vartheta_{\mathcal{R}}) - \Delta\ell\, h_{+|1}(0)\right) \frac{(\Delta y^2 - \Delta z^2)}{2} + \nonumber \\ & \hspace{25pt}\epsilon_{\times} \left(m_{\times|1}(\vartheta_{\mathcal{R}}) - \Delta\ell\, h_{\times|1}(0)\right) \Delta y \Delta z  \bigg]  \bigg\}. \label{delta}
\end{align}

Particularizing Eqs.~(\ref{t})--(\ref{z}) to ray 1 and replacing on them Eqs.~(\ref{A1}), (\ref{B1}) and (\ref{delta}), the general parametric equations for ray 1 are found:
\begin{widetext}
	\begin{align}
	\xi^t_{|1}(\vartheta) = t_{\mathcal{E}} & + \frac{\vartheta {\omega_{\textrm{e}}}_{\mathcal{E}}}{\Delta \ell} \bigg\{\Delta \ell - \frac{1}{(\Delta\ell - \Delta x)} \bigg[\epsilon_{\times} \Delta y \Delta z h_{\times|1}(0) + \epsilon_+ \frac{(\Delta y^2 - \Delta z^2)}{2} h_{+|1}(0)  \bigg]\bigg\} \nonumber \\ & + \frac{1}{(\Delta\ell - \Delta x)^2}  \bigg[\epsilon_{\times} m_{\times|1}(\vartheta) \Delta y \Delta z  + \epsilon_+ m_{+|1}(\vartheta) \frac{(\Delta y^2 - \Delta z^2)}{2}\bigg]  .\label{t_ray_1} 
	\end{align}	
	\begin{align}
	\xi^x_{|1}(\vartheta) &=  {x_{}}_{\mathcal{S}}  +  \frac{\vartheta {\omega_{\textrm{e}}}_{\mathcal{E}}}{\Delta\ell}\bigg\{\Delta x - \epsilon_+ \bigg[\frac{h_{+|1}(0) \Delta x}{\Delta\ell - \Delta x} + \frac{m_{+|1}(\vartheta_{\mathcal{R}})}{\Delta\ell}\bigg] \frac{(\Delta y^2 - \Delta z^2)}{2\Delta\ell}- \epsilon_{\times} \bigg[\frac{ h_{\times|1}(0) \Delta x}{\Delta\ell - \Delta x} + \frac{m_{\times|1}(\vartheta_{\mathcal{R}})}{\Delta\ell}\bigg] \frac{\Delta y \Delta z}{\Delta\ell} \bigg\} \nonumber \\ &\hspace{25pt}+\frac{1}{(\Delta\ell- \Delta x)^2} \bigg[\epsilon_+ m_{+|1}(\vartheta) \frac{\Delta y^2 - \Delta z^2}{2} + \epsilon_{\times} m_{\times|1} (\vartheta) \Delta y \Delta z  \bigg]\,,   \label{x_ray_1} 
	\end{align}
	\begin{align}
	\xi^y_{|1}(\vartheta) = {y_{}}_{\mathcal{S}} &+ \frac{\vartheta {\omega_{\textrm{e}}}_{\mathcal{E}}}{\Delta\ell}\bigg\{\Delta y  + \frac{\Delta y}{\Delta\ell(\Delta x - \Delta\ell)}   \left[\epsilon_{\times} \Delta y \Delta z  \, h_{\times|1}(0) + \epsilon_+ \frac{(\Delta y^2 - \Delta z^2)}{2}  h_{+|1}(0)\right] \nonumber \\ & \hspace{57pt}+\frac{\epsilon_{\times}m_{\times|1}(\vartheta_{\mathcal{R}})\Delta z}{\Delta\ell^2(\Delta x - \Delta\ell)^2} \left[\Delta y^2 \Delta x  + (\Delta x-\Delta \ell)(\Delta\ell^2-2\Delta y ^2)\right]\nonumber \\ &\hspace{57pt} + \frac{\epsilon_+ m_{+|1}(\vartheta_{\mathcal{R}}) \Delta y}{2\Delta\ell^2(\Delta x - \Delta\ell)^2}\left[ \Delta\ell^2 \Delta x + (\Delta x - 2\Delta\ell)  (\Delta x^2 + 2 \Delta z^2)\right] \hspace{-4pt}\bigg\} \nonumber \\ &+\frac{1}{\Delta\ell - \Delta x} \left[\epsilon_+ m_{+|1}(\vartheta) \Delta y + \epsilon_{\times} m_{\times|1}(\vartheta) \Delta z \right] ,\label{y_ray_1}  
	\end{align}
	\begin{align}
	\xi^z_{|1}(\vartheta) = {z_{}}_{\mathcal{S}} &+  \frac{\vartheta {\omega_{\textrm{e}}}_{\mathcal{E}}}{\Delta\ell}\bigg\{\Delta z + \frac{\Delta z}{\Delta\ell(\Delta x - \Delta\ell)} \left[\epsilon_{\times} \Delta y \Delta z  \, h_{\times|1}(0) 
	+ \epsilon_+ \frac{(\Delta y^2 - \Delta z^2)}{2}  h_{+|1}(0) \right]  \nonumber \\ & \hspace{57pt} - \frac{\epsilon_{\times}m_{\times|1}(\vartheta_{\mathcal{R}})\Delta y}{\Delta\ell^2(\Delta x - \Delta\ell)^2}\left[\Delta z^2 \Delta x  + (\Delta x-\Delta \ell)(\Delta\ell^2-2\Delta z^2)\right]  \nonumber\\& \hspace{57pt}  + \frac{\epsilon_+ m_{+|1} (\vartheta_{\mathcal{R}}) \Delta z}{2\Delta\ell^2(\Delta x - \Delta\ell)^2}\left[ \Delta\ell^2 \Delta x + (\Delta x - 2\Delta\ell)  (\Delta x^2 + 2 \Delta y^2)\right] \hspace{-4pt}\bigg\} \nonumber\\& +\frac{1}{\Delta\ell - \Delta x}  \left[\epsilon_{\times} m_{\times|1}(\vartheta) \Delta y - \epsilon_+ m_{+|1}(\vartheta) \Delta z\right] ,\label{z_ray_1} 
	\end{align}
\end{widetext}
The equations for ray 2 are obtained by exchanging ${x{}}^i_{\mathcal{S}} $ and $x^i_{\mathcal{M}}$, so that $\Delta x^i \rightarrow -\Delta x^i$; additionally, we also change $m_{P|1}(\vartheta_{\mathcal{R}}) \rightarrow m_{P|2}(\vartheta_{\mathcal{D}})$, $h_{P|1}(0) \rightarrow h_{P|2}(0)$ and ${\omega_{\textrm{e}}}_{\mathcal{E}}\rightarrow \omega_{\textrm{e}|2}(0) = \omega_{\textrm{e}|1}(\vartheta_{\mathcal{R}})$ on the above equations ($\omega_{\textrm{e}|1}(\vartheta_{\mathcal{R}})$ can be obtained later from Eq.~($\ref{omega1}$)). 

With the value of $\delta_{|1}$, $\vartheta_{\mathcal{R}}$ can also be determined. Evaluating Eq.~(\ref{u}) at $\vartheta_{\mathcal{R}}$, using Eqs.~(\ref{tR-tE}) and (\ref{sametime}), together with (\ref{delta}):
\begin{align}
&\vartheta_{\mathcal{R}} = \frac{\Delta x - \Delta t_{\mathcal{E,R}}}{\delta_{|1}} = \frac{1}{\omega_{e\mathcal{E}}} \bigg\{ \Delta \ell + \nonumber \\ &\frac{1}{(\Delta \ell - \Delta x)} \bigg[\epsilon_{\times} K_{\times} \Delta y \Delta z + \epsilon_+ K_+ \frac{\Delta y^2 -  \Delta z^2}{2}\hspace{-1pt} \bigg]   \hspace{-3pt}\bigg\}. \label{eq:vartheta_R}
\end{align}
where, remembering Eqs.~($\ref{M_P_1}$) and ($\ref{M_P_2}$), 
\begin{align}
\epsilon_P K_P \leqdef \,& \epsilon_P \left(\frac{\Delta x - 2 \Delta \ell}{\Delta \ell(\Delta \ell - \Delta x)}m_{P|1}(\vartheta_{\mathcal{R}(\bm{0})}) + h_{P|1}(0)\right) \nonumber \\ = \,& \epsilon_P\bigg(\frac{\Delta x - 2 \Delta \ell}{\Delta \ell(\Delta \ell - \Delta x)}M_{P|1}(t_{\mathcal{D}} - \Delta \ell) +  \nonumber \\ & \hspace{80pt} h_{P}(t_{\mathcal{D}} - 2\Delta \ell - x_{\mathcal{S}}) \bigg).
\end{align}
Making the already mentioned changes to calculate the back-trip analogue of $\vartheta_{\mathcal{R}}$, we find
\begin{align}
&\vartheta_{\mathcal{D}} = \frac{1}{\omega_{e\mathcal{E}}} \bigg[ \Delta \ell +\bigg(\epsilon_{\times} Q_{\times} \Delta y \Delta z + \epsilon_+ Q_+ \frac{\Delta y^2 -  \Delta z^2}{2}\hspace{-1pt} \bigg)   \hspace{-3pt}\bigg], \label{eq:vartheta_D}
\end{align}
where
\begin{align}
\epsilon_P Q_P \leqdef \, & \epsilon_P \bigg(\frac{h_{P|2}({\vartheta_{\mathcal{R}}}_{(\bm{0})})}{\Delta \ell + \Delta x} - \frac{\Delta h_{P|1}}{\Delta \ell - \Delta x}  \nonumber \\  &\hspace{40pt} -\frac{\Delta x + 2 \Delta \ell}{\Delta \ell(\Delta \ell + \Delta x)^2}m_P({\vartheta_{\mathcal{D}}}_{(\bm{0})}) \bigg)
\nonumber \\ = \,& \epsilon_P \bigg(\frac{h_{P}(t_{\mathcal{D}} - \Delta \ell - x_{\mathcal{M}})}{\Delta \ell + \Delta x} - \frac{\Delta h_{P|1}}{\Delta \ell - \Delta x} \nonumber \\ &\hspace{30pt}- \frac{\Delta x + 2 \Delta \ell}{\Delta \ell(\Delta \ell + \Delta x)^2}M_{P|2}(t_{\mathcal{D}} - \Delta \ell) \bigg)
\end{align}
with $\Delta h_{P|1} \leqdef h_{P|1}(\vartheta_{\mathcal{R}}) - h_{P|1}(0)$.
These expressions will be useful in calculating the final intensity pattern in a GW detector in section \ref{subsec:finalintensity}. A further discussion about the domains of the parametrized curves and their split in unperturbed and perturbed parts is held in Appendix \ref{app:domains}.

\subsection{TT frame kinematics and the optical parameters} \label{subsec:kin_parameters}

A reference frame, conceived as a congruence of observers \cite{Sachs1977}, has its local kinematics characterized by the irreducible decomposition:
\begin{equation}
\nabla_{\beta} u_{\alpha} = - a_{\alpha} u_{\beta} + \frac{1}{3}\Theta p_{\alpha \beta} + \sigma_{\alpha \beta} + \Omega_{\alpha \beta}\,, \label{irrdecomp}
\end{equation}
where $p_{\alpha \beta} \leqdef g_{\alpha \beta} + u_{\alpha} u_{\beta}$ is the projector onto the local rest space. The expansion scalar $\Theta$, the shear tensor $\sigma_{\alpha \beta}$ and the vorticity tensor $\Omega_{\alpha \beta}$ describe the local relative motion among observers of the frame. Together with the 4-acceleration $a^{\alpha}$, we refer to them as the kinematic quantities (or parameters) of the reference frame $u^\alpha$ \cite{Ellis2012, Ehlers1961, Ehlers1993}.

For the metric in Eq.~(\ref{metric}) and the frame of Eq.~(\ref{TTframe}) we find, from Eq.~($\ref{gamma^t}$),  
\begin{equation}
\nabla_{\beta} u_{\alpha} = \Gamma^t_{\beta \alpha} = \frac{1}{2}\epsilon_P h^P_{\beta \alpha,t} = \sigma_{\alpha \beta}, \label{eq:nablau}
\end{equation}
where
\begin{equation}
\sigma_{\alpha \beta} \leqdef p^{\gamma}_{(\beta} p^{\delta}_{\alpha)} \nabla_{\gamma} u_{\delta} - \frac{\Theta p_{\alpha \beta}}{3}, 
\label{Sigma}
\end{equation}
from which we conclude that the TT frame has a purely shearing kinematics induced by the presence of GWs. The usual heuristic infinitesimal (arm length much smaller than GW wavelength) description regarding interferometer response to a GW is closely attached to this kinematic property of the TT frame, to which the interferometer is assumed to be adapted. The detector deforms in an anisotropic way, such that the area of the rectangle defined by having the beam splitter and both of the end mirrors as its vertices is preserved. This is the expected non-trivial contribution to the RHS of Eq.~(\ref{eq:electric_evolution}), even when the interferometer is not in the long wavelength limit.

One may also define the so-called optical quantities $\widehat{a}^{\alpha}$, $\widehat{\Theta}$, $\widehat{\sigma}_{\alpha \beta}$  and $\widehat{\Omega}_{\alpha \beta}$, which are the kinematic parameters analogues for the description of a null congruence of curves \cite{Ellis2012} and are obtained by changing $u^{\alpha}$ for $k^{\alpha}$ and $p^{\alpha}_{\;\beta}$ for  the projector onto the local screen space: the 2-dimensional space orthogonal to $u^{\alpha}$ and to the spatial direction along which light propagates, $n^{\alpha} \leqdef p^{\alpha}_{\;\beta} k^{\beta}/(-u^{\mu} k_{\mu})$. 

Here we are concerned with the description of light rays,  working under the geometrical optics approximation (cf. Appendix \ref{app:geometrical_optics}) where, for a given scalar function $\psi$,
\begin{equation}
k_{\mu} = \nabla_{\mu} \psi\,, \label{phase}
\end{equation}
so that the null congruence is geodesic and irrotational ($\widehat{a}^{\alpha} = 0$ and $\widehat{\Omega}_{\alpha \beta} = 0$). Furthermore, the Faraday tensor in this regime reads \cite{Santana2020}:
\begin{equation}
F_{\mu \nu} = \frac{(k_{\mu} E_{\nu} - k_{\nu} E_{\mu})}{\omega_{\textrm{e}}}\,,
\end{equation}
where $E^{\mu} \leqdef F^{\mu \nu} u_{\nu}$ is the electric field measured by $u^{\mu}$. An equivalent assertion is that $F_{\mu \nu}$ is a null bivector \cite{Hall2004}, namely, it obeys:
\begin{equation}
F_{\mu \nu}F^{\mu \nu} = 0 = \frac{1}{2}\eta^{\alpha \beta \mu \nu} F_{\alpha \beta} F_{\mu \nu}\,, 
\label{nullfield}
\end{equation} 
implying, finally, that $\widehat{\sigma}_{\alpha \beta} = 0$, if $F^{\mu \nu}$ satisfies Maxwell equations in vacuum \cite{Robinson1961}. 

The optical expansion $\widehat{\Theta} \leqdef {k^{\mu}}_{;\mu}$ is then the only non-vanishing optical parameter when light rays travel in vacuum following the usual laws of electrodynamics. It gives the divergence ($\widehat{\Theta}>0$) or convergence ($\widehat{\Theta}<0$) property of a beam of light (cf. Eq.~(\ref{Thetaandarea})) and its evolution \cite{Ellis2012} along any of the rays simplifies to
\begin{equation}
\frac{d\widehat{\Theta}}{d \vartheta} + \frac{1}{2} \widehat{\Theta}^2 = 0 \label{Thetaevol}
\end{equation}  
when the spacetime is Ricci-flat and no electromagnetic 4-current is present. The above evolution law influence the electric field propagation, as depicted in Eq.~(\ref{eq:electric_evolution}), a subject addressed further in section \ref{sec:interferometry}.

\section{Frequency shift}
\label{sec:Doppler_effect}

The frequency shift of light in a GW spacetime was originally discussed in \cite{Kaufmann1970}, and is derived in \cite{Kopeikin1999} for a photon deflected with large impact parameter by a localized gravitational source. It is usually studied in the context of Doppler tracking of spacecrafts \cite{Estabrook1975, Tinto1998, Armstrong2006} and when discussing laser frequency noise in the LISA interferometer \cite{Tinto2002}. Here, our ultimate aim is to study this effect in a way that its contributions to the fluctuations in the final intensity pattern of an interferometric process can be properly identified. 

Starting from Eq.~(\ref{initfreq}), we will derive in subsection \ref{subsec:radar_doppler_effect}, for a non-monochromatic plane GW, the electromagnetic frequency evolution along the rays and consequently its total variation after a complete round-trip of light between $\mathcal{S}$ and $\mathcal{M}$. We interpret its origin in terms of the radar distance as the usual Doppler effect.

In addition, in subsection \ref{subsec:freq_vs_radar_dist} we address one of the most frequent questions regarding the detection of GWs through interferometry: ``If a GW stretches the arm and the laser wavelength simultaneously, should not these effects cancel each other? How can we detect GWs then?". In \cite{Saulson1997}, a qualitative discussion is made without the use of an explicit mathematical formalism. In \cite{Faraoni2007}, a quantitative discussion is made, but only in the long GW wavelength limit. In this work, we study this question relying only on the geometrical optics laws as the mathematical framework for light propagation.

\subsection{Doppler effect} \label{subsec:radar_doppler_effect}

Differentiating Eq.~(\ref{t_ray_1}) with respect to $\vartheta$, with the help of Eqs.~(\ref{np_deriv}) and (\ref{delta}), the frequency of light along ray 1 is obtained.
\begin{align}
\omega_{\textrm{e}|1}(\vartheta) =&\, {\omega_{\textrm{e}}}_{\mathcal{E}} \bigg\{1 + \frac{1}{\Delta\ell(\Delta\ell - \Delta x)} \times \nonumber \\ & \bigg[\epsilon_+ [h_{+|1}(\vartheta) - h_{+|1}(0)]  \frac{(\Delta y^2 - \Delta z^2)}{2}+  \nonumber \\ &\hspace{5pt} \epsilon_{\times} [h_{\times|1}(\vartheta)  - h_{\times|1}(0)]   \Delta y \Delta z \bigg]\bigg\}. \label{omega1}	
\end{align}
For ray 2, a similar expression is derived, by replacing $\omega_{e\mathcal{E}}$ for $\omega_{e|2}(0) = \omega_{e|1}(\vartheta_{\mathcal{R}})$ and making the change $\Delta x^i \rightarrow - \Delta x^i$. Evaluating the resulting expression in $\vartheta_{\mathcal{D}}$ we arrive at the frequency shift measured by observer $\mathcal{S}$ after a round-trip of light
\begin{widetext}
\begin{align}
\frac{\Delta \omega_{\textrm{e}}}{{\omega_{\textrm{e}}}_{\mathcal{E}}} (t_{\mathcal{D}}) \leqdef \,  \frac{\omega_{\textrm{e}}(\vartheta_{\mathcal{D}}) - {\omega_{\textrm{e}}}_{\mathcal{E}}}{{\omega_{\textrm{e}}}_{\mathcal{E}}}= &\big \{\epsilon_+ [h_+(t_{\mathcal{D}} - x_{\mathcal{S}}) - h_+(t_{\mathcal{D}} - \Delta\ell - x_{\mathcal{M}})] \cos{2\phi} \, \nonumber \\ & + \epsilon_{\times}[h_{\times}(t_{\mathcal{D}} - x_{\mathcal{S}}) - h_{\times}(t_{\mathcal{D}} - \Delta\ell - x_{\mathcal{M}})] \sin{2\phi} \big\} \sin^2{\Big(\frac{\theta}{2}\Big)} \nonumber \\  & + \big\{\epsilon_+[h_+(t_{\mathcal{D}} - \Delta\ell - x_{\mathcal{M}})  -h_+(t_{\mathcal{D}} - 2\Delta\ell - x_{\mathcal{S}})]  \cos{2\phi}\, \nonumber \\ &  + \epsilon_{\times}[ h_{\times}(t_{\mathcal{D}} - \Delta\ell - x_{\mathcal{M}}) -h_{\times}(t_{\mathcal{D}} - 2\Delta\ell - x_{\mathcal{S}})] \sin{2\phi}\big\} \cos^2{\Big(\frac{\theta}{2}\Big)} , \label{redshift}
\end{align}
\end{widetext}
where we have used Eq.~(\ref{sph}) to express it in terms of $(\theta,\phi)$. Note that both Eq.~(\ref{radardistspheric}) and Eq.(\ref{redshift}) are undetermined when the GW is propagating parallel to the photons ($\theta=0$). This is not a problem, since to derive these expressions we have assumed $\delta \neq 0$ and this indeterminacy occurs for $\delta =0$. In this special case, the radar distance is not perturbed by GWs and there is no frequency shift as well, once null geodesics coincide with those of the flat background.

This expression is in agreement with the ones presented in works like \cite{Armstrong2006, Estabrook1975}. It is the percentage difference either between initial and final frequency of light in a round-trip, or between the perturbed and unperturbed frequency at the final event.

From the purely shearing kinematics of the TT frame induced by the GWs (cf. \ref{subsec:kin_parameters}), to which observers $\mathcal{S}$ and $\mathcal{M}$ belong, that manifests itself by the radar distance change with time, there is a clear indication that the above Doppler effect should exist. Indeed, as pointed out in e.g. \cite{Ellis1971}, the frequency evolution along a ray, in general, obeys
\begin{equation}
	\frac{1}{\omega_e}\frac{d \omega_e}{d \vartheta} = - \left(\frac{1}{3} \Theta + a_{\mu}n^{\mu}+\sigma_{\mu\nu}n^{\mu}n^{\nu}\right)\omega_e\,. \label{eq:freq_kin_orig}
\end{equation}
In fact, the change in radar distance and the frequency shift are actually two facets of the same GW-photon interaction. 
By differentiating Eq.~(\ref{D_R}) with respect to $t_{\mathcal{Q}}$ and having Eqs.~(\ref{M_P_1}), (\ref{M_P_2}) and (\ref{np_deriv}) in mind, one concludes that
\begin{equation}	
\frac{d D_{R,\xi}}{d t_{\mathcal{Q}}} (t_{\mathcal{Q}}(t_{\mathcal{D}})) = - \frac{1}{2} \frac{\Delta \omega_{\textrm{e}}}{\omega_{\textrm{e}\mathcal{E}}}(t_{\mathcal{D}}). \label{Doppler}
\end{equation}
This is indeed the usual expression describing a Doppler effect when relative velocities are small compared to the speed of light in vacuum, if we interpret the covariant radar velocity defined by
\begin{equation}
v_{R} \leqdef  \frac{d D_{R,\xi}}{dt_{\mathcal{Q}}}  \label{radar_velocity}
\end{equation}
as a relative velocity of observer $\mathcal{M}$ with respect to $\mathcal{S}$. Relation ($\ref{Doppler}$) can be trivially obtained in any spacetime by taking the ratio between the proper time differences of two emitted wave crests and the subsequently received ones \cite{Schutz2009}; the factor 1/2 arises from the accumulation of Doppler effects in the two rays of the round-trip.

In a general context, where the distance between two observers is not small compared to the GW wavelength, their relative motion cannot be described in terms of geodesic deviations and, moreover, in this non-local framework, a relative velocity cannot be uniquely defined by means of a comparison between velocities at different events, given the curved character of spacetime. On the other hand, Eqs.~(\ref{Doppler}) and (\ref{radar_velocity}) highlight how suitable is the covariant description of the relative motion between observers $\mathcal{M}$ and $\mathcal{S}$ in terms of the radar distance, regardless of how far they are from each other, without the need to compare their 4-velocities, since $v_R$ is measured solely in terms of information obtained by $\mathcal{S}$.

\begin{figure*}[t]
	\includegraphics[scale=0.5]{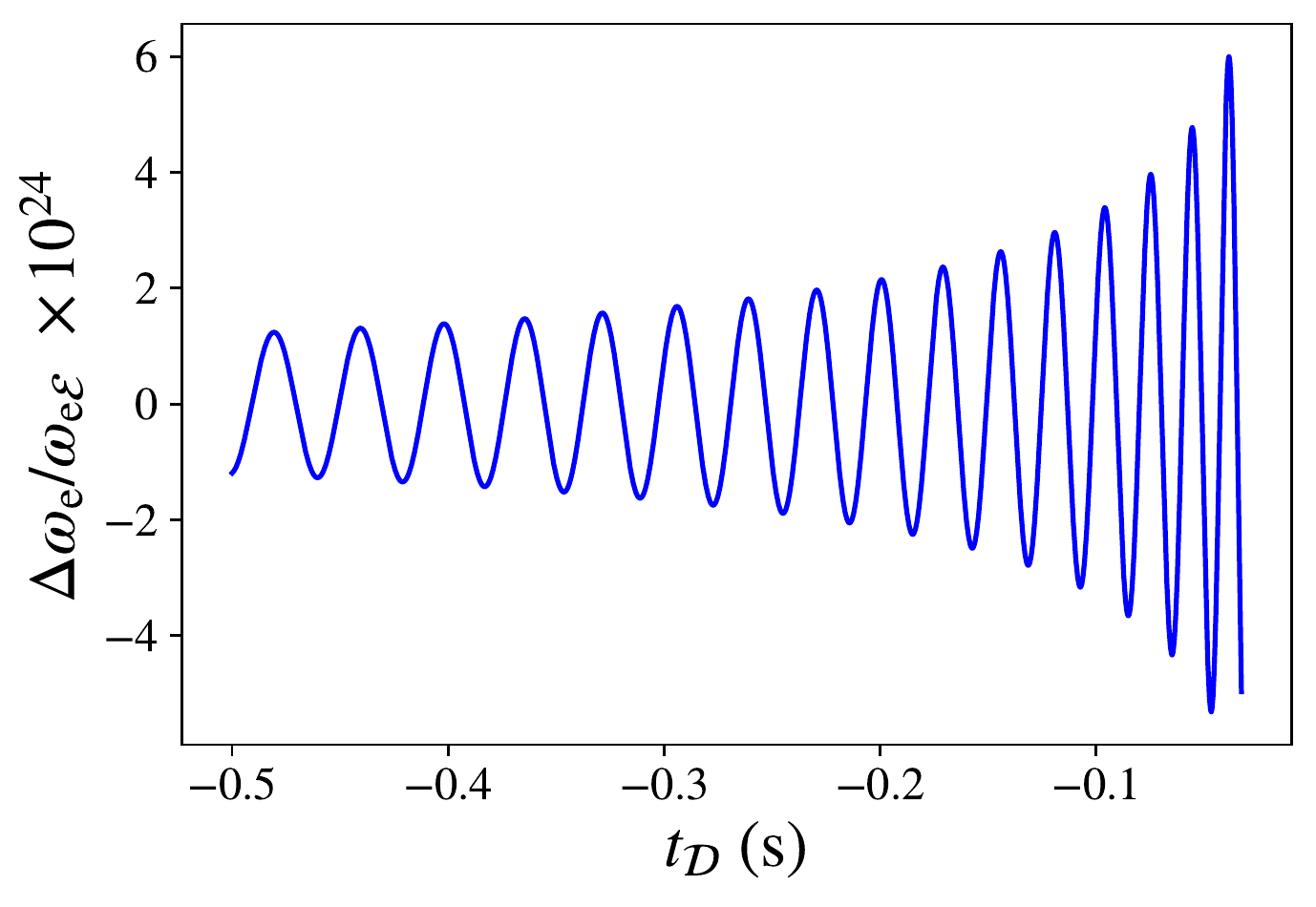} \hspace{30pt}
	\includegraphics[scale=0.5]{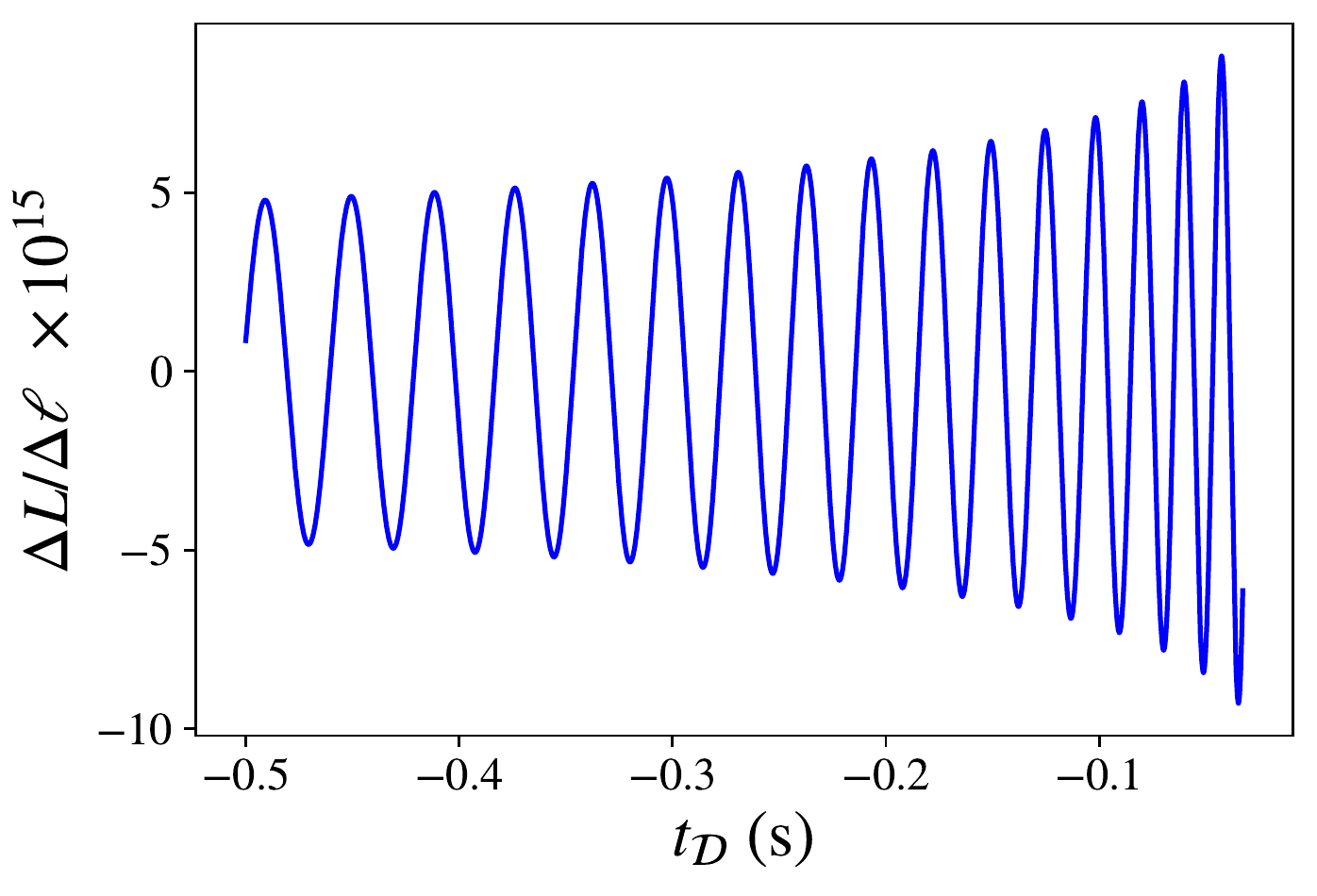}
	\caption{Frequency shift (left) and radar length perturbation (right) for a typical waveform of a binary system detected by aLIGO in the inspiral phase (up to a frequency $\omega_g \simeq$ 425 rad/s). We have made $\theta = \pi/2$, $\phi = \pi/3$, $\Delta \ell =$ 4 km, $M_c = 28.3 \, \text{M}_{\odot}$, $r = 410$ Mpc and $\Phi_0=t_c=\iota = 0$.}
	\label{fig:binary_waveform}
\end{figure*}

Finally, for illustrative purposes, Fig.~\ref{fig:binary_waveform} shows two key quantities explicitly computed until this point, namely, light frequency shift in Eq.~(\ref{redshift}) and the radar distance perturbation
\begin{align}
\Delta L \leqdef D_R(t_{\mathcal{D}} - \Delta \ell) - \Delta \ell \label{eq:Delta_L_def}
\end{align}
in Eq.~(\ref{radardistspheric}), for a simple template of the GW amplitude emitted by a binary merger in its inspiral phase. It is given by \cite{Maggiore2007}:
\begin{align}
h_+(t-x) &= \left(\frac{GM_c}{c^2}\right)^{5/4} \left(\frac{5}{c \tau}\right)^{1/4} \frac{1 + \cos^2 \iota}{2r} \cos[\Phi(\tau)]\,,\nonumber \\
h_{\times}(t-x) &= \left(\frac{GM_c}{c^2}\right)^{5/4} \left(\frac{5}{c \tau}\right)^{1/4} \frac{\cos \iota}{r} \sin[\Phi(\tau)]\,, \label{eq:gw_binary}
\end{align}
where $M_c$ is the chirp mass of the binary system, $r$ is the luminosity distance from the source to the detector, $\iota$ is the angle between the normal direction to the binary plane and the line of sight and
\begin{equation}
\Phi(\tau) \leqdef -2 \left(\frac{5GM_c}{c^3} \right)^{-5/8} \tau^{5/8} + \Phi_0\,,
\end{equation}
with
\begin{equation}
\tau \leqdef t_{c} - t - x/c\,,
\end{equation}
where $\Phi_0$ is a constant and $t_{c}$ is the instant of coalescence of the merger as seen by an observer at $x = 0$. The values of $M_c$ and $r$ were chosen to match those of the first detected black hole binary system by aLIGO \cite{Abbott2016b}.

\subsection{Can frequency shift and arm length perturbation combine to cancel the interference pattern?} \label{subsec:freq_vs_radar_dist}

In interferometric experiments designed to detect GWs, one usually relies on the assumption that the linearized perturbation of the final intensity pattern is directly related with the phase difference between light beams after traveling their round-trip along each arm. If one assumes that the amplitude and polarization of the electromagnetic fields of the beams are not affected by the GW, this relation follows and is widely adopted in literature (e.g. \cite{Maggiore2007}). For this subsection, we shall seek this point of view to discuss one of the most common conceptual concerns arising as a consequence of this picture. However, as we shall demonstrate in subsection (\ref{subsec:finalintensity}), there are additional contributions to the interference pattern if the electromagnetic field is evolved according to its full propagation equation along null geodesics in curved spacetimes \cite{Santana2020}. 

The issue usually posed \cite{Saulson1997, Faraoni2007} is based on the following question: should not the frequency shift in Eq.~(\ref{redshift}) result in a contribution to the final phase difference, \emph{additionally} to that related to the difference in round-trip travel times derived from Eq.~(\ref{radardistspheric})? After all, the change in light wavelength should heuristically result in a change of the distances between the electric field maxima. If this is true, could such contribution cancel the first in some specific case, so that no GW could be detected at all?

In the LIGO FAQ webpage \cite{LIGOFAQ}, the answer to the intimately related question ``If a gravitational wave stretches the distance between the LIGO mirrors, doesn't it also stretch the wavelength of the laser light?" begins with the following assertion: ``While it's true that a gravitational wave does stretch and squeeze the wavelength of the light in the arms \emph{ever so slightly}, it does NOT affect the fact that the beams will travel different distances as the wave changes each arm's length". But why is it that the only relevant quantity for the difference in phase is the difference in path traveled by light is never justified in the answer. 

Furthermore, at the end, the webpage concludes by stating: ``The effects of the length changes in the arms far outweigh any change in the wavelength of the laser, so we can virtually ignore it altogether.". Firstly, these effects, though minute, are of the same order in $\bm{\epsilon}$, preventing any \emph{a priori} conclusions about the negligibility of any of them when compared with the other. Although in the \emph{particular} case studied in Fig.~\ref{fig:binary_waveform} the change in the arms' lengths indeed outweighs the change in the wavelength, this assertion does not seem straightforward for other regimes in the GW spectrum. In \cite{Faraoni2007}, Faraoni shows for the long-wavelength limit ($\lambdabar_g \gg \Delta\ell$, where $\lambdabar_g\leqdef \lambda_g/2\pi$ is the reduced GW wavelength) that, in fact, the frequency shift vanishes. In this frequency band the conclusion in the webpage is then justified. But for the higher portion of the aLIGO detectable range of frequencies ($1$ kHz to $10$ kHz), such approximation breaks down (for $1$ kHz, $\lambdabar_g \approx 10$ km, while $\Delta\ell = 4$ km). Besides, for detectors like LISA and the Cosmic Explorer, this approximation becomes even less applicable, since the former has $\Delta\ell = 2.5 \times 10^6$ km and will detect GWs in the range $10^{4} \text{\;km}< \lambdabar_g < 10^8 \text{\;km}$, and the latter has $\Delta\ell = 40$ km, detecting in the range $5 \text{\;km}<\lambdabar_g<10^4 \text{\;km}$. Lastly, even if it were true that one effect is always dominant over the other, it is not obvious how do they contribute to the final intensity pattern so that one could be neglected when compared to the other. Here we provide a simple answer to the problem raised regardless of any assumption on either the GW wavelength or the arm length, provided that the geometrical optics approximation for light is valid.    

In the electromagnetic geometrical optics regime, the phase of light  $\psi(x^{\mu})$ is related to the 4-dimensional wave vector by Eq.~(\ref{phase}). This equation suggests, since $\omega_{\textrm{e}} = k_t$ for the TT frame, that $\psi$ is affected by how the frequency evolves along the light ray. However, the nullity of light geodesic rays (\ref{pert_geod}) together with Eq.~(\ref{phase}) gives:
\begin{equation}
	k_{\mu}k^{\mu} = 0 \Rightarrow \frac{D \psi}{d \vartheta} = 0. \label{constphase}
\end{equation} 
This, in turn, implies that, although it is true that there is a frequency shift influence on the phase throughout the photon's round-trip, the changes in the spatial components $k_{i}$ or, in other words, the spatial trajectory perturbations, contribute to it as well, in such a way that the net effect is to preserve the constancy of $\psi$ throughout the null curve. 

One can conclude from these considerations that the phase at the end of the round-trip of the beams in each arm is equal to its initial value. But, as portrayed on the second panel of Fig.~(\ref{fig:toymodel}), because the interferometer's arms are deformed distinctly, the rays that combine at the end must have been emitted in different events along $\mathcal{S}$. The initial phase depends only on the initial frequency and the emission time $t_{\mathcal{E}} = t_{\mathcal{D}} - 2 D_R$, which, for a given detection time depends only on the radar length of arm traveled, and so, in general, the rays have different values of $\psi$. Consequently, the phase difference at the end occurs, indeed, solely because of the discrepant paths light travels in each arm. If one assumes this phase disparity to be the only contribution to the linearized intensity fluctuations, frequency shifts, although existent, should not explicitly stand as an independent or additional influence to be considered in the intensity measurements, but to be another manifestation of the change in radar distance, as illustrated by Eq.~(\ref{Doppler}). Nonetheless, we shall demonstrate in subsection (\ref{subsec:finalintensity}) that if the electric field magnitude is evolved using its full propagation equation \cite{Santana2020}, a frequency shift contribution does arise from the perturbation on each photon's energy. 
%


\section{Electric field evolution and interference pattern}
\label{sec:interferometry}

\begin{figure}[t]
	\centering
	\includegraphics[width=0.23\textwidth]{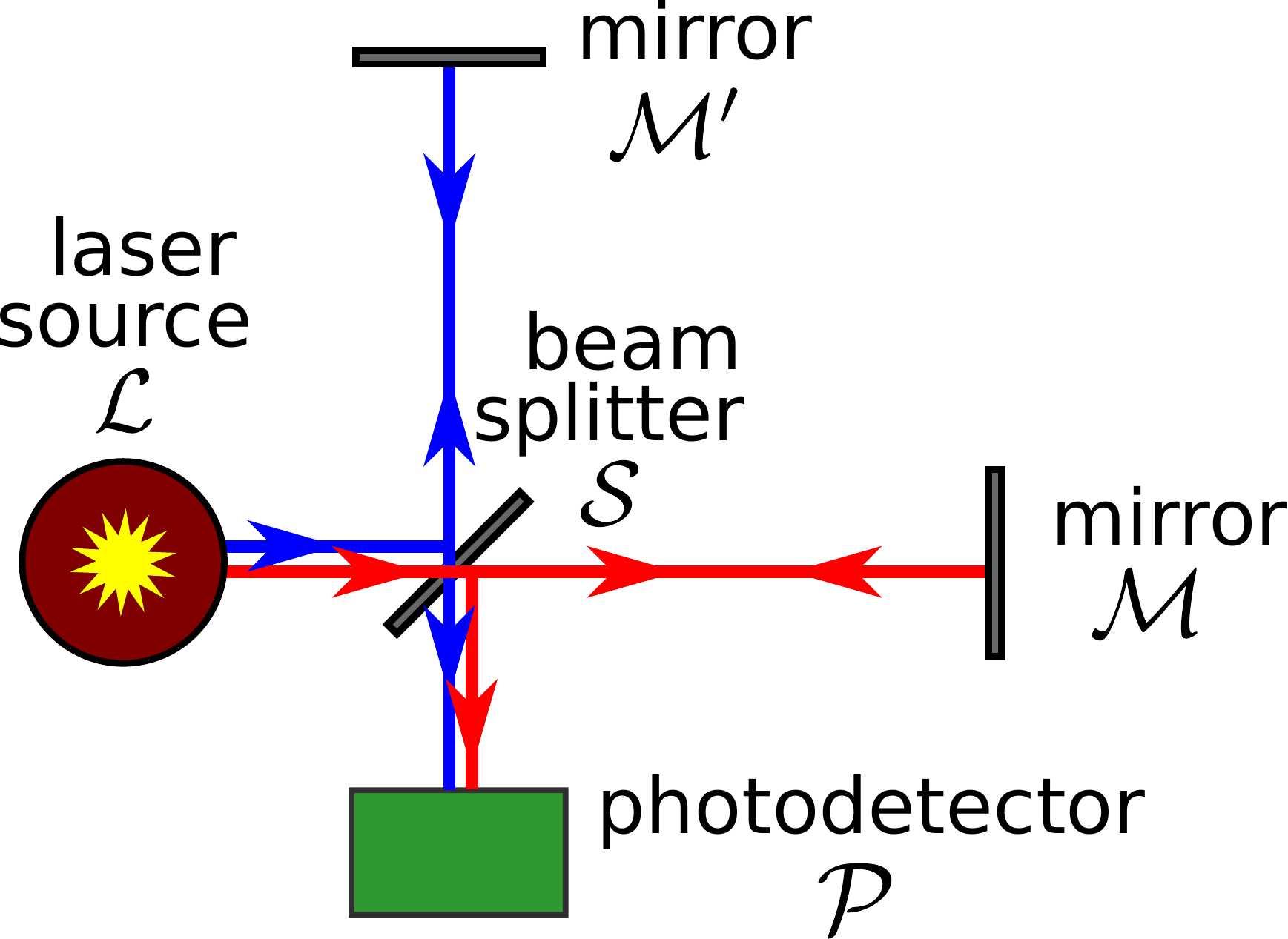} \hspace{5pt}
	\includegraphics[width=0.23\textwidth]{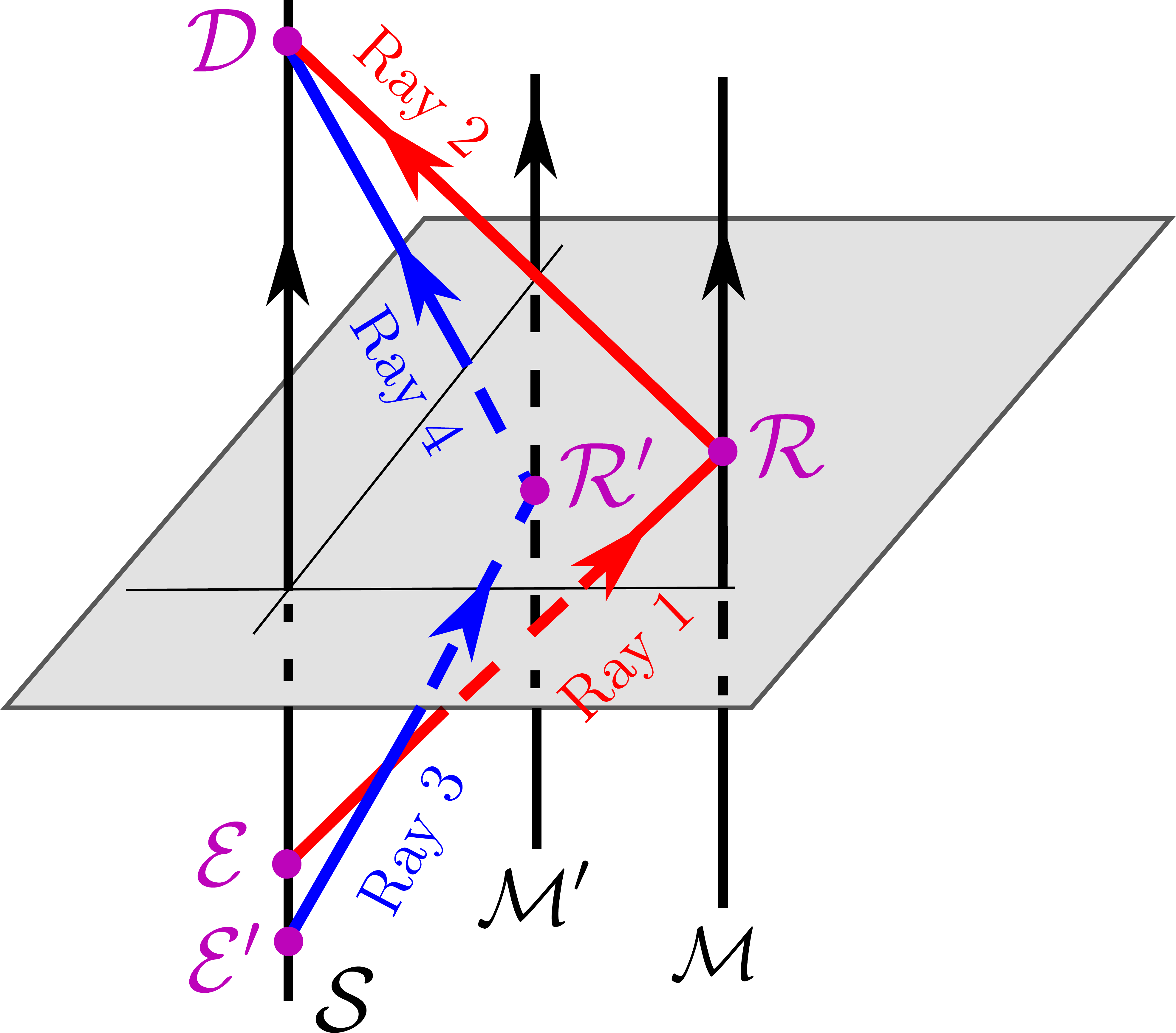}
	\caption{\emph{Left}: Spatial diagram of a Michelson-Morley GW interferometer. \emph{Right}: Corresponding spacetime diagram, with relevant worldlines: tilted ones for the null geodesic arcs and vertical ones, for the several devices. Although the above images illustrates perpendicular arms, we here assume general arms orientations.}
	\label{fig:toymodel}
\end{figure}

In this section, we will consider a toy model for a Michelson-Morley interferometry experiment used to detect GWs. A pictorial description of such a model is represented on the first panel of Fig.~\ref{fig:toymodel}\;. There, light is emitted from a laser $\mathcal{L}$ to a beam splitter $\mathcal{S}$ which separates the original signal into two equal-power halves. These, in turn, propagate to the end mirrors $\mathcal{M}$ and $\mathcal{M}'$ of their correspondent arms, where they are reflected back to the beam splitter and finally recombine at the photo-detector $\mathcal{P}$, providing a time-dependent interference pattern for light. 

This is a toy model due to several aspects. First, most of the technological elements of a real GW detector are neglected here, in particular, the usual Fabry-P\'{e}rot cavities, since interesting aspects are already apparent without them and the calculations are much simplified. Second, the distances between $\mathcal{L}$ and $\mathcal{S}$ and between $\mathcal{S}$ and $\mathcal{P}$ will be neglected, since these are small when compared with the arm's length. With this last observation in mind, the second panel of Fig.~\ref{fig:toymodel} expresses the 4-dimensional description of the experiment, with observer $\mathcal{S}$ sending, in different instants, the beams that will interfere at the final event on this same observer. A third idealized characteristic of our model is the point-like nature of the mirrors. In this case, for an interferometry experiment to be successful, the light rays must be sent in a very restrictive way from $\mathcal{S}$ to each of the mirrors. To that end, the boundary conditions specified by the data from Eq.~(\ref{mixed_conditions}) must be imposed on the light rays traveling each arm as we did in previous sections when finding the null geodesic arcs exchanged by two observers. Therefore, several results previously derived regarding such arcs will stand as valid and important ingredients to our approach of the interferometric process, being only necessary to change $\mathcal{M}$ for $\mathcal{M'}$ when dealing with the other arm. As already mentioned, a fourth and final aspect of the idealization worth emphasizing is the assumption that light propagates as a test field, not generating any relevant curvature despite being affected by the presence of GWs.

We shall assume that the TT observers $\mathcal{S}$, $\mathcal{M}$ and $\mathcal{M}'$ have arbitrary spatial coordinates $x^i_{\mathcal{S}}$, $x^i_{\mathcal{M}}$ and $x^i_{\mathcal{M}'}$, respectively, and consequently the arms are not necessarily orthogonal. Furthermore, as indicated on the second panel of Fig.~\ref{fig:toymodel}, the relevant rays and events of the $\mathcal{S}$$\mathcal{M}$ arm are denoted as in our previous sections, while the emission and reflection events for the $\mathcal{S}$$\mathcal{M'}$ arm will be $\mathcal{E}'$ and $\mathcal{R}'$, respectively, and rays 3 and 4 will denote outgoing and incoming rays. Primed symbols will generically stand for quantities relative to the $\mathcal{S}$$\mathcal{M'}$ arm. Our ultimate goal is to obtain the final interference pattern, and, thus, we shall propagate, using Eq.~(\ref{eq:electric_evolution}) and the metric of Eq.~(\ref{metric}), the electric field as measured by the observers of Eq. (\ref{TTframe}) along rays 1, 2, 3 and 4 up to the common detection event $\mathcal{D}$.

\subsection{Solving for the optical expansion along a ray} \label{optical_expansion}

We start by calculating the remaining ingredient for solving Eq.~(\ref{eq:electric_evolution}), namely, the optical expansion $\widehat{\Theta}$ along a chosen light ray (cf. subsection (\ref{subsec:kin_parameters})). From Eq.~(\ref{Thetaevol}):
\begin{equation}
	\widehat{\Theta}_{|j}(\vartheta) = \frac{2\widehat{\Theta}_{|j}(0)}{2 + \vartheta \widehat{\Theta}_{|j}(0)}\,, \label{Thetasol}
\end{equation}
for $j = 1, 2, 3, 4$. We notice that, choosing $\vartheta > 0$ along the curve, if initially the beam is divergent, it will remain in this way, as one would expect, since Eq.~(\ref{Thetaevol}) is also valid in Minkowski spacetime. Assuming a divergent beam from now on, we find the integrating factor of Eq.~(\ref{eq:electric_evolution}) for ray 1:
\begin{equation}
	\textrm{e}^{-\frac{1}{2}\int_{0}^{\vartheta_{\mathcal{R}}} \widehat{\Theta}_{|1} d\vartheta} = \frac{1}{1+\frac{\vartheta_{\mathcal{R}}\widehat{\Theta}_{\mathcal{E}} }{2}}, \label{eq:exptheta1}
\end{equation}
where $\widehat{\Theta}_{\mathcal{E}} \leqdef \widehat{\Theta}_{|1}(0)$. A similar expression is valid for ray 2, being only necessary to make $\vartheta_{\mathcal{R}} \rightarrow \vartheta_{\mathcal{D}}$, $\widehat{\Theta}_{|1} \rightarrow \widehat{\Theta}_{|2}$ and $\widehat{\Theta}_{\mathcal{E}} \rightarrow  \widehat{\Theta}_{|2}(0)$.

Here we shall restrict to the case in which the optical expansion is continuously connected from ray 1 to ray 2, namely,
\begin{equation}
	\widehat{\Theta}_{|2}(0) = \widehat{\Theta}_{|1}(\vartheta_{\mathcal{R}}). \label{contTheta}
\end{equation}
Imposing this initial condition for ray 2 in Eq.~(\ref{Thetasol}), 
\begin{equation}
	\textrm{e}^{-\frac{1}{2}\int_{0}^{\vartheta_{\mathcal{D}}} \widehat{\Theta}_{|2} d\vartheta} = \frac{1}{1+\frac{\vartheta_{\mathcal{D}}\widehat{\Theta}_{\mathcal{E}} }{2+\vartheta_{\mathcal{R}} \widehat{\Theta}_{\mathcal{E}}}}\,. \label{eq:exptheta2final}
\end{equation}

Note that this continuity assumption is only valid once a passive reflection in event $\mathcal{R}$ is guaranteed by means of a plane mirror. If the mirror was concave, $\widehat{\Theta}$ would have an abrupt change after reflection from a positive to a negative value; if it was convex, the discontinuity would still occur, although $\widehat{\Theta}$ would remain positive. The passive reflection can be contrasted with what occurs in a LISA-like interferometer \cite{Danzmann2017}. Since the arms in LISA are designed to be huge, the intensity of the emitted light dissipates and only a small amount reaches the other extremity. It is then necessary to generate a new, but phase-locked light beam at event $\mathcal{R}$ so that the final intensity pattern can be substantial in magnitude. Condition (\ref{contTheta}) cannot be achieved in such a framework, where one would more naturally have to impose
$\widehat{\Theta}_{|2}(0) = \widehat{\Theta}_{|1}(0)\,$.

To find $\widehat{\Theta}_{\mathcal{E}}$, we first relate the optical expansion with the beam's cross section area \cite{Ellis2012} (notice the typo in Eq.~(7.25) therein):
\begin{equation}
	\widehat{\Theta} = \frac{1}{\delta S}\frac{d}{d\vartheta}(\delta S)\,. \label{Thetaandarea}
\end{equation}
On the aLIGO experiment, the beam containing ray 1 has, to zeroth order $\bm{\epsilon}$, initial and final cross section areas given by \cite{Aasi2015, Martynov2016}:
\begin{align}
	\delta S(0) & = 88 \,\text{cm}^2\,, \label{initarea} \\ 
	\delta S(\vartheta_{\mathcal{R}}) & = 121 \,\text{cm}^2\,. \label{finalarea}
\end{align} 
Because of the small dimension of the laser, we will assume that the initial condition for the optical expansion is not affected by GWs, \emph{i.e} it does not depend on $\bm{\epsilon}$. Integrating Eq.~(\ref{Thetaandarea}) from $0$ to $\vartheta_{\mathcal{R}}$, inserting Eqs.~(\ref{initarea}), (\ref{finalarea}) and evaluating the LHS integral by Eq.~(\ref{Thetasol}) one concludes that:
\begin{equation}
	\widehat{\Theta}_{\mathcal{E}} =  \frac{0.345}{{\vartheta_{\mathcal{R}}}_{(\bm{0})}} = 0.52 \times 10^3 \; \text{m}^{-2}, \label{eq:init_Theta_est}
\end{equation}
since, from Eq.~(\ref{eq:vartheta_R}), ${\vartheta_{\mathcal{R}}}_{(\bm{0})} = c \Delta \ell/\omega_{e \mathcal{E}}$, where $\Delta \ell = 4$ km and $\omega_{e \mathcal{E}} = 1.8 \times 10^{15}$ rad/s  \cite{Martynov2016}. Eq.~(\ref{eq:init_Theta_est}) will be useful in estimating final interference pattern contributions related to intensity dissipation due to the beam divergence on the aLIGO case.

Of course the procedure developed in this subsection can be trivially extended to the other arm by changing ray 1 to 3 and 2 to 4 and the corresponding emission and reflection events (see Fig.~\ref{fig:toymodel}). Also, our modeling of the experiment assumes hereafter that the initial frequency and optical expansion remain the same throughout the emission events along $\mathcal{S}$, in particular, although $\mathcal{E'} \neq \mathcal{E}$
\begin{align}
	\omega_{e \mathcal{E'}} = \omega_{e \mathcal{E}}\,, \quad
	\widehat{\Theta}_{\mathcal{E'}} = 	\widehat{\Theta}_{\mathcal{E}}\,.
\end{align}

\subsection{The electric field evolution in the geometric optics limit} \label{subsec:electriceq}
Here we reserve a space to discuss the physical significance of Eq.~(\ref{eq:electric_evolution}) for the propagation of electric fields along light rays  under the geometrical optics limit on arbitrary spacetimes when these are measured by a general reference frame whose field of instantaneous observers is given by $u^{\mu}$. 

Expressing $E^{\mu} \reqdef E e^{\mu}$, where $E \leqdef \sqrt{E^{\mu} E_{\mu}}$, we may dismember such equation into two parts \cite{Santana2020}:
\begin{equation}
	\frac{De^{\mu}}{d\vartheta} = k^{\mu} \frac{e^{\nu}}{\omega_{\textrm{e}}} \frac{Du_{\nu}}{d\vartheta}\,,
	\label{polevol}
\end{equation}
and
\begin{equation}
	\frac{dE}{d\vartheta} + \frac{\widehat{\Theta}}{2}E = -\frac{k^{\nu}E}{\omega_{\textrm{e}}}\frac{Du_{\nu}}{d\vartheta} = \frac{E}{\omega_{\textrm{e}}}\frac{d\omega_{\textrm{e}}}{d\vartheta}\,. \label{Magevol}
\end{equation}

From Eq.~(\ref{polevol}), one concludes that the first term of the RHS of Eq.~(\ref{eq:electric_evolution}) is present if and only if the electric field polarization $e^{\mu}$ is not parallel transported along the light ray. For a Faraday tensor satisfying Eq.~(\ref{nullfield}) the (instantaneous) intensity of light is given by
\begin{equation}
	I = g_{\mu\nu} E^\mu E^\nu\,,
\end{equation}
and from Eq.~(\ref{Magevol}) it is easy to deduce how it propagates, namely
\begin{equation}
	\frac{dI}{d\vartheta} + \widehat{\Theta} I = \frac{2I}{\omega_{\textrm{e}}}\frac{d\omega_{\textrm{e}}}{d\vartheta}\,,
	\label{eq:intensity_evolution}
\end{equation}
from which, together with Eq.~(\ref{Thetaandarea}), we obtain
\begin{equation}
	\frac{I\delta S}{\omega_{\textrm{e}}^2} = \text{const}
	\label{eq:brightness_conservation}
\end{equation}
along the chosen geodesic, which stands for the conservation of photon number of the light beam \cite{Schneider1992}.

The intimate relation between Eqs.~(\ref{Magevol}) and (\ref{eq:intensity_evolution}) allows us to interpret the physical origins of the other terms present in Eq.~(\ref{eq:electric_evolution}). In Eq.~(\ref{Magevol}), the RHS of each equality demonstrates that the second term on the RHS of Eq.~(\ref{eq:electric_evolution}) is a consequence of a frequency shift effect arising from, as attested by Eq.~(\ref{eq:freq_kin_orig}), the kinematics of the reference frame. The physical interpretation surrounding it is clear once one notices that the energy of each photon is proportional to its frequency and that, as a result, a shift in $\omega_{\textrm{e}}$ should alter the intensity of light, which is confirmed by Eq.~(\ref{eq:brightness_conservation}). Finally, the term proportional to $\widehat{\Theta}$ accounts for the increase or decrease in intensity following a possible convergence or divergence of the beam, respectively.

In order to compute $Du^{\mu}/d\vartheta$ in Eq.(\ref{eq:electric_evolution}), although it is only necessary to define a differentiable vector field of instantaneous observers on the light ray of interest, in our case one needs an entire reference frame if the electric field is to be propagated along all possible rays continuously emitted for all possible interferometer orientations. In this case, in view of Eq.~(\ref{eq:nablau}), such an absolute derivative can be expressed,  in terms of the kinematics of the TT frame:
\begin{equation}
	\frac{Du^{\mu}}{d\vartheta} = k^{\nu} {\sigma^{\mu}}_{\nu}\,.
\end{equation}   
As a consequence, we highlight that it is the shearing character of this frame that induces the non-parallel transport of the polarization vector and the frequency shift of light, which contribute to a non-trivial propagation of $E^{\mu}$.

One could try to use directly Eq.~(\ref{eq:intensity_evolution}) to propagate the intensity in each ray and calculate the final interference pattern of the experiment. But, since we allow the polarization vector of rays 2 and 4 on event $\mathcal{D}$ to be differently perturbed by the GW, the relation between the final intensity on each of these two rays before interference with the total intensity after it is not, in general, obvious. This is why we choose to solve Eq.~(\ref{eq:electric_evolution}) instead.  
\subsection{Electric field solution} \label{subsec:electricsolutions}

We begin by noting that the non-trivial part of Eq.~(\ref{eq:electric_evolution}) is the spatial one, since $g_{ti} = 0$ and, thus
\begin{equation}
	E^t \leqdef F^{t \nu} u_{\nu} = 0\,.
\end{equation} 
Taking into account Eq.~(\ref{eq:nablau}) and that, as presented in appendix \ref{app:christoffel_symbols}, all Christoffel symbols are of order $\bm{\epsilon}$, it can be conveniently rewritten as:
\begin{equation}
	\frac{dE^{i}}{d\vartheta} + \frac{1}{2} \widehat{\Theta} E^{i} = f^i, \label{eq:electricpropGW}
\end{equation}
with
\begin{align}
	f^i(\vartheta) \leqdef k^{\beta}_{(\bm{0})} \bigg[&\bigg(  \frac{\zeta k^{i}_{(\bm{0})} E^{\nu}_{(\bm{0})}(\vartheta) - k^{\nu}_{(\bm{0})} E^{i}_{(\bm{0})}(\vartheta)}{\omega_{e(\bm{0})}}\bigg) \Gamma^t_{\beta \nu} (\vartheta)   \nonumber \\ &- \Gamma^{i}_{\beta j} (\vartheta) E^j_{(\bm{0})}(\vartheta) \bigg], \label{eq:f}
\end{align}
where the last term arises from the absolute derivative of the electric field. The quantity $\zeta = 0,1$ was introduced by hand to monitor the presence of the contribution related to the RHS of Eq.~(\ref{polevol}), allowing us to conclude whether the non-parallel transport of $e^{\mu}$ induced by the TT frame kinematics will result in additional contributions to the electric field along each ray.

Such a propagation equation is then solved perturbatively as follows. To zeroth order, Eq.~(\ref{eq:electricpropGW}) becomes
\begin{equation}
	\frac{dE^{i}_{(\bm{0})}}{d\vartheta} + \frac{1}{2} \widehat{\Theta}_{(\bm{0})}E^{i}_{(\bm{0})} = 0,
\end{equation}
whose solution is simply
\begin{equation}
	E^i_{(\bm{0})}(\vartheta) = E^i_{(\bm{0})}(0) \textrm{e}^{-\frac{1}{2}\int_{0}^{\vartheta}\hat{\Theta}_{(\bm{0})}(\vartheta')d\vartheta'}. \label{eq:zerothorder}
\end{equation}
As for the linear order solution, one can easily integrate Eq.~(\ref{eq:electricpropGW}) with the help of an integrating factor. Once Eq.~(\ref{eq:zerothorder}) is replaced on such a solution, one finds: 
\begin{align}
	&E^i(\vartheta) = \textrm{e}^{-\frac{1}{2}\int_{0}^{\vartheta}\hat{\Theta}(\vartheta')d\vartheta'} \bigg\{E^i(0) + \nonumber \\ &\hspace{15pt} k^{\beta}_{(\bm{0})}  \bigg[\frac{[\zeta k^{i}_{(\bm{0})} E^{\nu}_{(\bm{0})}(0) - k^{\nu}_{(\bm{0})} E^{i}_{(\bm{0})}(0)]}{\omega_{\textrm{e}\mathcal{E}}}  \int_{0}^{\vartheta} \Gamma^t_{\beta \nu} (\vartheta') d\vartheta'  \nonumber \\ & \hspace{35pt}- E^{j}_{(\bm{0})}(0) \int_{0}^{\vartheta} \Gamma^{i}_{\beta j}(\vartheta') d\vartheta' \bigg]   \bigg\}\,, \label{eq:electricmidway}
\end{align}
where we note that the unperturbed electromagnetic frequency along the ray $\omega_{\textrm{e}(\bm{0})}$ equals its initial value $\omega_{\textrm{e}\mathcal{E}}$.
The first integral in Eq.~(\ref{eq:electricmidway}), together with Eqs.~(\ref{gamma^t}) and (\ref{u}), can then be solved:
\begin{align}
	\int_{0}^{\vartheta} \Gamma^{t}_{\beta \nu}(\vartheta') d\vartheta' & = 	\frac{\epsilon_P}{4}\int_{0}^{\vartheta} h^{P}_{\beta \nu,u}(\vartheta') d\vartheta'  \nonumber \\ & = \frac{\epsilon_P}{2}\int_{0}^{\vartheta} \frac{dh^{P}_{\beta \nu}}{d\vartheta'} \bigg(\frac{dw}{d\vartheta'}\bigg)^{-1} d\vartheta' \nonumber \\ & =-\frac{\epsilon_P}{2 \delta_{(\bm{0})}} [h^P_{\beta \nu}(\vartheta) - h^P_{\beta \nu}(0)]\,. \label{eq:inth}
\end{align}
The remaining integrals are analogously handled with the help of Eqs.~($\ref{gamma_t}$)--($\ref{gamma_z}$).

We now particularize to ray 1. The corresponding null geodesics in Minkowski spacetime obey:
\begin{equation}
	k^{\alpha}_{\bm{(0)}|1} = \omega_{\textrm{e}\mathcal{E}} \bigg(\delta^{\alpha}_t + \frac{\Delta x^i}{\Delta \ell} \delta^{\alpha}_i \bigg). \label{eq:kMink}
\end{equation}
Evaluating Eq.~($\ref{eq:electricmidway}$) on ray 1, replacing the solved Christoffel symbol integrals together with Eqs.~(\ref{delta}) and (\ref{eq:kMink}), and recalling Eqs.~(\ref{hplus}) and (\ref{hcross}), one concludes that: 
\begin{equation}
	E^i_{|1}(\vartheta) = \textrm{e}^{-\frac{1}{2}\int_{0}^{\vartheta}\widehat{\Theta}_{|1}(\vartheta')d\vartheta'} [E^i_{|1}(0) + J^i_{|1}(\vartheta)]\,, \label{eq:Eray1}
\end{equation}
where:
\begin{widetext}
\begin{align}
	& \hspace{-50pt} J^x_{|1}(\vartheta) = \frac{1}{2(\Delta \ell-\Delta x)}  \bigg\{ \epsilon_+ \Delta h_{+|1}(\vartheta) \bigg[\bigg( 1 - \zeta \frac{\Delta x}{\Delta \ell}\bigg) \big(\Delta y E^y_{(\bm{0})|1}(0) -\Delta z E^z_{(\bm{0})|1}(0) \big) + \frac{E^x_{(\bm{0})|1}(0)}{\Delta \ell}\big(\Delta y^2 -\Delta z^2\big) \bigg]  \nonumber \\ & \hspace{-50pt} \hspace{89pt}+ \epsilon_{\times} \Delta h_{\times|1}(\vartheta) \bigg[ \bigg( 1 - \zeta \frac{\Delta x}{\Delta \ell}  \bigg) (\Delta y E^z_{(\bm{0})|1}(0) + \Delta z E^y_{(\bm{0})|1}(0)) + 2\frac{\Delta z \Delta y}{\Delta \ell}E^x_{(\bm{0})|1}(0) \bigg]\bigg\}, \label{M1}
\end{align}
\begin{align}
	&\hspace{-4pt}J^y_{|1}(\vartheta) = \frac{1}{2(\Delta \ell-\Delta x)} \bigg\{ \epsilon_+ \Delta h_{+|1}(\vartheta)\bigg[ E^z_{(\bm{0})|1}(0)\frac{\zeta \Delta z \Delta y}{\Delta \ell} - E^x_{(\bm{0})|1}(0) \Delta y  + E^y_{(\bm{0})|1}(0)\bigg(\Delta \ell - \Delta x + \frac{\Delta y^2(1 - \zeta) - \Delta z^2}{\Delta \ell}\bigg) \bigg] \nonumber \\ & \hspace{-4pt} \hspace{89pt} + \epsilon_{\times} \Delta  h_{\times|1}(\vartheta)\bigg[ E^y_{(\bm{0})|1}(0)\frac{\Delta y \Delta z}{\Delta \ell}(2 - \zeta) - E^x_{(\bm{0})|1}(0)\Delta z  + E^z_{(\bm{0})|1}(0)\bigg( \Delta \ell - \Delta x - \frac{\zeta \Delta y^2}{\Delta \ell} \bigg) \bigg]\bigg\}\label{M2},
\end{align}
\begin{align}
	&\hspace{-4pt}J^z_{|1}(\vartheta) = \frac{1}{2(\Delta \ell-\Delta x)}\bigg\{ \epsilon_+\Delta h_{+|1}(\vartheta)\bigg[E^x_{(\bm{0})|1}(0)\Delta z  - E^y_{(\bm{0})|1}(0)\frac{\zeta \Delta z \Delta y}{\Delta \ell} + E^z_{(\bm{0})|1}(0)\bigg( \Delta x - \Delta \ell + \frac{\Delta z^2(\zeta - 1) + \Delta y^2}{\Delta \ell}\bigg)  \bigg] \nonumber \\
	& \hspace{85pt}+ \epsilon_{\times}\Delta h_{\times|1}(\vartheta)\bigg[ E^z_{(\bm{0})|1}(0)\frac{\Delta y \Delta z}{\Delta \ell}(2 - \zeta) - E^x_{(\bm{0})|1}(0) \Delta y + E^y_{(\bm{0})|1}(0)\bigg( \Delta \ell - \Delta x - \frac{\zeta \Delta z^2}{\Delta \ell} \bigg) \bigg]\bigg\}. \label{eq:M3}
\end{align}
\end{widetext}
and $\Delta h_{P|j}(\vartheta) \leqdef h_{P|j}(\vartheta) - h_{P|j}(0)$. 

With the aid of Fig.~\ref{fig:toymodel}, it is easy to see that the field on ray 3 is obtained from Eqs.~(\ref{eq:Eray1})--(\ref{eq:M3}) by changing $\Delta x^i \rightarrow \Delta x'^{i} \leqdef x_{\mathcal{M'}} - x_{\mathcal{S}}$, $\Delta \ell \rightarrow \Delta \ell' \leqdef \sqrt{\delta_{ij}\Delta x'^{i}\Delta x'^{j}}$, $\widehat{\Theta}_{|1} \rightarrow \widehat{\Theta}_{|3}$, $E^i_{|1}(0) \rightarrow  E^i_{|3}(0)$ and $\Delta h_{P|1} \rightarrow \Delta h_{P|3}$. On the other hand, the field on ray 2 is a consequence of making the changes $\Delta x^i \rightarrow -\Delta x^{i}$, $\widehat{\Theta}_{|1} \rightarrow \widehat{\Theta}_{|2}$, $\Delta h_{P|1} \rightarrow \Delta h_{P|2}$ and $E^i_{|1}(0) \rightarrow E^i_{|2}(0) = -E^i_{|1}(\vartheta_{\mathcal{R}})$; the latter change expressing the usual phase shift of $\pi$ when light is assumed to be reflected by a perfect mirror. Finally, for ray 4, one should do the same changes when passing from ray 1 to 2, but on the expression of $E^i_{|3}(\vartheta)$.

To calculate the final intensity pattern on our toy model interferometer, the fields coming from rays 2 and 4 must be evaluated at event $\mathcal{D}$ (cf. Fig.~\ref{fig:toymodel}). From the above results, $E^i_{|2}(\vartheta_{\mathcal{D}})$ can be recast in terms of the initial value of the electric field $E^i_{|1}(0)$, the GW amplitude $h_P$ and the known quantities $\Delta x^i$, $\omega_{\textrm{e}\mathcal{E}}$ and $\widehat{\Theta}_{\mathcal{E}}$. To that end, it suffices to notice that, from the analogue of Eq.~(\ref{eq:Eray1}) for ray 2
\begin{align}
E^i_{|2}(\vartheta_{\mathcal{D}}) &= \textrm{e}^{-\frac{1}{2}\int_{0}^{\vartheta_{\mathcal{D}}}\widehat{\Theta}_{|2}(\vartheta')d\vartheta'} [-E^i_{|1}(\vartheta_{\mathcal{R}}) + J^i_{|2}(\vartheta_{\mathcal{D}})] \nonumber \\&= \textrm{e}^{-\frac{1}{2}\int_{0}^{\vartheta_{\mathcal{D}}}\widehat{\Theta}_{|2}(\vartheta')d\vartheta'} \Big\{-\textrm{e}^{-\frac{1}{2}\int_{0}^{\vartheta_{\mathcal{R}}}\widehat{\Theta}_{|1}(\vartheta')d\vartheta'} \times \nonumber \\ & \hspace{30pt} \times [E^i_{|1}(0)+J_{|1}^i(\vartheta_{\mathcal{R}})] + J^i_{|2}(\vartheta_{\mathcal{D}})\Big\}, \label{eq:Eray2}
\end{align}
where, in $J^i_{|2}(\vartheta_{\mathcal{D}})$, the initial field $E^i_{(\bm{0})|2}(0)$ can be expressed in terms of $E^i_{(\bm{0})|1}(0)$ as
\begin{equation}
E^i_{(\bm{0})|2}(0) = - E^i_{(\bm{0})|1}(0) \textrm{e}^{-\frac{1}{2}\int_{0}^{\vartheta_{\mathcal{R}}}\widehat{\Theta}_{|1}(\vartheta')d\vartheta'}, \label{eq:initE2}
\end{equation}
so that the exponential terms can be factored out. Then, it is only necessary to replace Eqs.~(\ref{eq:exptheta1}) and (\ref{eq:exptheta2final}) on Eq.~(\ref{eq:Eray2}) together with Eqs.(\ref{eq:vartheta_R}) and (\ref{eq:vartheta_D}).
\subsection{Intensity pattern for GW normal incidence} \label{subsec:finalintensity}
 Here, to simplify our analysis, we restrict ourselves to the particular case in which the incidence of the GW (which is traveling in the $x$ direction) is normal to the detector apparatus, namely, we consider $\Delta x = \Delta x' = 0$. 

Let an auxiliary ray, call it ray 5, be the one leaving event $\mathcal{E}$ together with ray 1, but on the other arm of the interferometer. These rays are the transmitted and reflected halves of the original laser beam, divided by the splitter. Immediately after this splitting, our starting event $\mathcal{E}$, their electric fields differ only by a minus sign due to the reflection
\begin{equation}
	E^i_{|5}(0) = - E^i_{|1}(0)\,, \label{eq:refl_phase_shift}
\end{equation}
and, thus, imposing the geometrical optics result of the transversality of the electric field, in zeroth order, on both rays at their common emission, we have
\begin{equation}
	k_{i (\bm{0})|1}(0)E^{i}_{(\bm{0})|1}(0) = - k_{i (\bm{0})|5}(0)E^{i}_{(\bm{0})|1}(0) = 0\,. \label{eq:transv_k_E}
\end{equation}
One concludes, therefore, in the case of normal GW incidence, that
\begin{equation}
	E^y_{(\bm{0})|1}(0) = E^z_{(\bm{0})|1}(0) = 0\,. \label{eq:vanishing_unpert_comp}
\end{equation}
An analogous argument is made to conclude that $E^y_{(\bm{0})|3}(0) = E^z_{(\bm{0})|3}(0) = 0\,.$

Here it is important to stress that such a result does not depend on the value of the angle between the arms of the interferometer, which is treated as arbitrary throughout this work. It is only a consequence of the fact that, at zeroth order, the initial electric field must be orthogonal to the rays simultaneously leaving the beam splitter in both arms which define a plane that, in the normal incidence case, is orthogonal to the GW propagation direction, namely the $yz$ plane. 
We emphasize that, in a more general physical situation, a general light ray whose spatial trajectory resides on the $yz$ plane does not necessarily have its zeroth order electric field parallel to the GW. That is the case for interferometers because of the beam splitter device, that connects the initial electric fields at each arm (Eq.~($\ref{eq:refl_phase_shift}$)). Finally, because of Eq.~($\ref{eq:zerothorder}$),
\begin{equation}
	E^y_{(\bm{0})|j}(\vartheta) = E^z_{(\bm{0})|j}(\vartheta)=0. \label{eq:zeroth_electric_field_vanish}
\end{equation}

With all of this in mind, $J^i_{|1}$ and $J^i_{|2}$ simplify to
\begin{align}
	J^x_{|j}(\vartheta) & = \frac{1}{2\Delta \ell^2} [\epsilon_+ \Delta h_{+|j}(\vartheta)(\Delta y^2 - \Delta z^2) + \nonumber \\ & \hspace{42pt} 2\epsilon_{\times} \Delta h_{\times|j}(\vartheta) \Delta y \Delta z] E^x_{(\bm{0})|j}(0), \label{eq:Jxnorm} \\
	J^y_{|j}(\vartheta) &= \frac{(-1)^j}{2 \Delta \ell} [\epsilon_+ \Delta h_{+|j}(\vartheta)\Delta y \,+ \nonumber \\ & \hspace{61pt} \epsilon_{\times} \Delta h_{\times|j}(\vartheta) \Delta z ] E^x_{(\bm{0})|j}(0), \label{eq:Jynorm} \\
	J^z_{|j}(\vartheta) &= \frac{(-1)^{j+1}}{2\Delta \ell} [\epsilon_+ \Delta h_{+|j}(\vartheta)\Delta z \,- \nonumber \\ & \hspace{61pt} \epsilon_{\times} \Delta h_{\times|j}(\vartheta) \Delta y ] E^x_{(\bm{0})|j}(0), \label{eq:Jznorm}
\end{align}
where $j=1,2$ in the above expressions. The exponential factors become 
\begin{align}
	& \textrm{e}^{-\frac{1}{2}\left[\int_{0}^{\vartheta_{\mathcal{D}}}\widehat{\Theta}_{|2}(\vartheta')d\vartheta' +\int_{0}^{\vartheta_{\mathcal{R}}}\widehat{\Theta}_{|1}(\vartheta')d\vartheta'\right]} = \frac{1}{1+\frac{\widehat{\Theta}_{\mathcal{E}}(\vartheta_{\mathcal{R}}+ \vartheta_{\mathcal{D}})}{2}}  \nonumber \\ &= \frac{\omega_{\textrm{e}\mathcal{E}}}{\widehat{\Theta}_{\mathcal{E}}\Delta \ell + \omega_{\textrm{e}\mathcal{E}}} \bigg \{1 - \frac{\widehat{\Theta}_{\mathcal{E}}}{(\widehat{\Theta}_{\mathcal{E}}\Delta \ell + \omega_{\textrm{e}\mathcal{E}})}  \bigg[2\Delta L(t_{\mathcal{D}}- \Delta \ell) \nonumber \\ &\hspace{70pt}+ \epsilon_{+} \frac{\Delta y^2 - \Delta z^2}{2\Delta \ell} h_{+}(t_{\mathcal{D}} - 2 \Delta \ell - x_{\mathcal{S}}) \nonumber \\ & \hspace{70pt}+ \epsilon_{\times}\frac{\Delta y \Delta z}{\Delta \ell} h_{\times}(t_{\mathcal{D}} - 2 \Delta \ell - x_{\mathcal{S}})\bigg]\bigg\}, \label{eq:expnorm}
\end{align}
where the radar distance perturbation of Eq.~(\ref{eq:Delta_L_def}), for normal incidence, simplifies to: 
\begin{align}
	\Delta L (t) &=  -\frac{1}{2} \bigg[\epsilon_+ \frac{\Delta y^2 - \Delta z^2}{2\Delta \ell^2} \int^{t+\Delta \ell -x_{\mathcal{S}}}_{t-\Delta \ell -x_{\mathcal{S}}} h_+(w) dw \nonumber \\ &\hspace{43pt}+ \epsilon_{\times} \frac{\Delta y \Delta z}{\Delta \ell^2} \int^{t+\Delta \ell -x_{\mathcal{S}}}_{t-\Delta \ell -x_{\mathcal{S}}} h_{\times}(w) dw \bigg]. \label{eq:Delta_L}
\end{align}

Inserting Eq.~($\ref{eq:initE2}$) in Eqs.~($\ref{eq:Jxnorm}$)--($\ref{eq:Jznorm}$) and the result in Eq.~($\ref{eq:Eray2}$) together with Eq.~($\ref{eq:expnorm}$), the final electric field on ray 2 reads:
\begin{align}
	E^x_{|2}(t_{\mathcal{D}}) & = - \frac{\omega_{\textrm{e}\mathcal{E}}E^x_{|1}(t_{\mathcal{E}})}{\widehat{\Theta}_{\mathcal{E}} \Delta \ell +  \omega_{\textrm{e}\mathcal{E}}} \bigg[1+ \frac{\Delta y^2 - \Delta z^2}{2\Delta \ell^2} \epsilon_+ F_+(t_{\mathcal{D}}) \nonumber \\& \hspace{87pt}+ \frac{\Delta y \Delta z}{\Delta \ell^2} \epsilon_{\times }F_{\times}(t_{\mathcal{D}})  \bigg], \label{eq: x_electr_comp_norm_inc}\\ \nonumber \\
	E^y_{|2}(t_{\mathcal{D}}) & = -\frac{\omega_{\textrm{e}\mathcal{E}}}{\widehat{\Theta}_{\mathcal{E}} \Delta \ell + \omega_{\textrm{e}\mathcal{E}}} \bigg\{E^y_{|1}(t_{\mathcal{E}}) +\nonumber \\ &\hspace{10pt}\frac{1}{2\Delta \ell}E^x_{(\bm{0})|1}(t_{\mathcal{E}})\bigg[\Delta y G_+(t_{\mathcal{D}}) + \Delta z G_{\times}(t_{\mathcal{D}})\bigg] \bigg\},\label{eq: y_electr_comp_norm_inc}\\ \nonumber \\
	E^z_{|2}(t_{\mathcal{D}}) & = \frac{\omega_{\textrm{e}\mathcal{E}}}{\widehat{\Theta}_{\mathcal{E}} \Delta \ell + \omega_{\textrm{e}\mathcal{E}}} \bigg\{E^z_{|1}(t_{\mathcal{E}}) + \nonumber \\ &\hspace{10pt} \frac{1}{2\Delta \ell}E^x_{(\bm{0})|1}(t_{\mathcal{E}})\bigg[\Delta z G_+(t_{\mathcal{D}}) - \Delta y G_{\times}(t_{\mathcal{D}})\bigg] \bigg\}, \label{eq: z_electr_comp_norm_inc}
\end{align}
where
\begin{align}
	F_{P}(t_{\mathcal{D}}) \leqdef & \, h_{P}(t_{\mathcal{D}} - x_{\mathcal{S}}) - h_{P}(t_{\mathcal{D}} - 2\Delta \ell - x_{\mathcal{S}})+ \nonumber\\ & \frac{\widehat{\Theta}_{\mathcal{E}}}{\widehat{\Theta}_{\mathcal{E}} \Delta \ell + \omega_{\textrm{e}\mathcal{E}}} \bigg[ \int^{t_{\mathcal{D}} -x_{\mathcal{S}}}_{t_{\mathcal{D}}-2\Delta \ell -x_{\mathcal{S}}} h_{P}(w) dw \nonumber \\&\hspace{60pt} -\Delta \ell h_{P}(t_{\mathcal{D}} - 2\Delta \ell - x_{\mathcal{S}}) \bigg]
\end{align}
and
\begin{align}
	G_P(t_{\mathcal{D}}) \leqdef\,& h_{P}(t_{\mathcal{D}} - x_{\mathcal{S}}) - h_{P}(t_{\mathcal{D}} - \Delta \ell - x_{\mathcal{M}}) - \nonumber \\ &  [h_{P}(t_{\mathcal{D}} - \Delta \ell - x_{\mathcal{M}}) - h_{P}(t_{\mathcal{D}} - 2\Delta \ell - x_{\mathcal{S}})].
\end{align}
Under the same circumstances, $E^i_{|4}$ may be obtained by making $\Delta y \rightarrow \Delta y'$, $\Delta z \rightarrow \Delta z'$, $x_{\mathcal{M}} \rightarrow x_{\mathcal{M'}}$, $\Delta \ell \rightarrow \Delta \ell'$ and $E^i_{|1}(t_{\mathcal{E}}) \rightarrow E^i_{|3}(t_{\mathcal{E'}})$ in Eqs.~($\ref{eq: x_electr_comp_norm_inc}$)--($\ref{eq: z_electr_comp_norm_inc}$). 

Here we notice the vanishing of the $\zeta$ contributions. This is expected in the case of normal incidence since, because of Eq.~ ($\ref{eq:zeroth_electric_field_vanish}$), the unperturbed polarization vector is parallel to the GW propagation direction, i.e. $e^{\nu}_{\bm{(0)}|j} = e^x_{\bm{(0)}|j} \delta^{\nu}_{x} , \forall j$, and so it is orthogonal to the shear tensor of Eq.~($\ref{Sigma}$), which implies Eq.~($\ref{polevol}$) to take the form:
\begin{equation}
	\frac{De^{\mu}_{|j}}{d\vartheta} =   k^{\mu}_{|j}k^{\alpha}_{|j} e^{\nu}_{|j} \sigma_{\alpha \nu}  = \frac{1}{2}\epsilon_P k^{\mu}_{|j}k^{\alpha}_{|j} e^{x}_{\bm{(0)}|j} h^P_{ \alpha x,t} = 0\,,
\end{equation}
and so the electric field polarization is indeed parallel transported. For arbitrary incidence, the shearing $yz$ plane will not coincide with the plane of the arms, allowing $e^{\nu}$ not to be perpendicular to $\sigma_{\alpha \beta}$, and so we expect $\zeta$ contributions not to vanish.


Then, computing the total electric field at $\mathcal{D}$
\begin{equation}
	E^{\mu}_T(t_{\mathcal{D}}) \leqdef E^{\mu}_{|2}(t_{\mathcal{D}}) +  E^{\mu}_{|4}(t_{\mathcal{D}}),
\end{equation} 
using Eqs.~(\ref{metric}) and (\ref{eq:zeroth_electric_field_vanish}), we find the final intensity:
\begin{align}
	I(t_{\mathcal{D}}) &= [\eta_{\mu \nu} + \epsilon_P h^P_{\mu \nu}(t_{\mathcal{D}} - x_{\mathcal{S}})] E^{\mu}_T(t_{\mathcal{D}}) E^{\nu}_T(t_{\mathcal{D}}) \nonumber \\ &= [E^x_T(t_{\mathcal{D}})]^2. \label{eq: final_intensity_simplification}
\end{align}

We then develop the usual procedure of relating the final electric fields to the distance traveled by each ray assuming a harmonic initial condition for the electric field:
\begin{align}
	E^x_{|1}(t_{\mathcal{E}}) &= \mathcal{E}^x \cos(\omega_{\textrm{e}\mathcal{E}}t_{\mathcal{E}}) \nonumber \\&= \mathcal{E}^x \cos[\omega_{\textrm{e}\mathcal{E}}(t_{\mathcal{D}} - 2D_{R}(t_{\mathcal{D}}- \Delta \ell))] \nonumber \\&= \mathcal{E}^x\{ \cos[\omega_{\textrm{e}\mathcal{E}}(t_{\mathcal{D}} - 2\Delta \ell)] + \nonumber \\ & \hspace{30pt}2 \omega_{e \mathcal{E}} \Delta L(t_{\mathcal{D}} - \Delta \ell) \sin[\omega_{e \mathcal{E}}(t_{\mathcal{D}} - 2\Delta \ell)]\}\,, \label{eq: x_comp_init_cond_ray_1}
\end{align}
where $\mathcal{E}^x$ is a constant amplitude along $\mathcal{S}$. For $E^x_{|3}(t_{\mathcal{E'}})$, we write:   
\begin{align}
	E^x_{|3}(t_{\mathcal{E'}}) &= -\mathcal{E}^x \cos(\omega_{\textrm{e}\mathcal{E}}t_{\mathcal{E'}})  \nonumber \\ &= -\mathcal{E}^x\{ \cos[\omega_{\textrm{e}\mathcal{E}}(t_{\mathcal{D}} - 2\Delta \ell'] + \nonumber \\&\hspace{30pt}2 \omega_{e \mathcal{E}} \Delta L'(t_{\mathcal{D}} - \Delta \ell') \sin[\omega_{e \mathcal{E}}(t_{\mathcal{D}} - 2\Delta \ell')]\}\,, \label{eq: x_comp_init_cond_ray_3}
\end{align}
where the relative negative sign is, again, a consequence of the initial reflection on the beam-splitter.

Replacing in Eq.~(\ref{eq: final_intensity_simplification}), Eq.~(\ref{eq: x_electr_comp_norm_inc}) and its equivalent expression for ray 4, together with Eqs.~(\ref{eq: x_comp_init_cond_ray_1}) and (\ref{eq: x_comp_init_cond_ray_3}), we are able to write (making $c\neq 1$) the interference pattern as a function of time:
\begin{widetext}
\begin{subequations} \label{eq:finalintensity}
	\begin{align}
	&I(t_{\mathcal{D}}) = I_{(\bm{0})}(t_{\mathcal{D}}) + \frac{2(\mathcal{E}^x)^2}{1 + c\widehat{\Theta}_{\mathcal{E}}\Delta \ell/\omega_{e\mathcal{E}}} \bigg\{\frac{\cos{[\omega_{\textrm{e}\mathcal{E}}(t_{\mathcal{D}} - 2\Delta \ell'/c)]}}{1 + c\widehat{\Theta}_{\mathcal{E}} \Delta \ell'/\omega_{\textrm{e}\mathcal{E}}} - \frac{\cos{[\omega_{\textrm{e}\mathcal{E}}(t_{\mathcal{D}} - 2\Delta \ell/c)]}}{1 + c\widehat{\Theta}_{\mathcal{E}} \Delta \ell/\omega_{\textrm{e}\mathcal{E}}}\bigg\} \left[T'(t_{\mathcal{D}}) - T(t_{\mathcal{D}})\right] \,,
	\end{align}
	with:
	\begin{align}
		I_{(\bm{0})}(t_{\mathcal{D}}) \leqdef (\mathcal{E}^x)^2 \bigg\{\frac{\cos{[\omega_{\textrm{e}\mathcal{E}}(t_{\mathcal{D}} - 2\Delta \ell'/c)]}}{1 + c\widehat{\Theta}_{\mathcal{E}} \Delta \ell'/\omega_{\textrm{e}\mathcal{E}}} - \frac{\cos{[\omega_{\textrm{e}\mathcal{E}}(t_{\mathcal{D}} - 2\Delta \ell/c)]}}{1 + c\widehat{\Theta}_{\mathcal{E}} \Delta \ell/\omega_{\textrm{e}\mathcal{E}}}\bigg\}^2\,,
	\end{align}
	and
	\begin{align}
		T(t_{\mathcal{D}}) \leqdef & \frac{2\omega_{\textrm{e}\mathcal{E}}}{c}\Delta L(ct_{\mathcal{D}} - \Delta \ell) \sin{[\omega_{\textrm{e}\mathcal{E}}(t_{\mathcal{D}} - 2 \Delta \ell/c)]} +  \frac{\Delta \omega_{\textrm{e}}}{\omega_{\textrm{e}\mathcal{E}}}(t_{\mathcal{D}}) \cos{[\omega_{\textrm{e}\mathcal{E}}(t_{\mathcal{D}} - 2\Delta \ell/c)]} \nonumber  \\ &- \frac{c\widehat{\Theta}_{\mathcal{E}}}{c\hat{\Theta}(t_{\mathcal{E}}) \Delta \ell + \omega_{\textrm{e}\mathcal{E}}} \cos{[\omega_{\textrm{e}\mathcal{E}}(t_{\mathcal{D}} - 2 \Delta \ell/c)]} \bigg[2 \Delta L(ct_{\mathcal{D}} - \Delta \ell)  \nonumber \\ &  + \frac{(\Delta y^2 - \Delta z^2)}{2\Delta \ell} \epsilon_+ h_{+}(ct_{\mathcal{D}} - 2\Delta \ell- x_{\mathcal{S}}) + \frac{\Delta y \Delta z}{\Delta \ell} \epsilon_{\times} h_{\times}(ct_{\mathcal{D}} - 2\Delta \ell - x_{\mathcal{S}}) \bigg] , \label{eq:non_trivial_pert}
	\end{align}
\end{subequations}
\end{widetext}
where $T'$ has the same expression but for the other arm, namely, with the changes $\Delta \ell \rightarrow \Delta \ell'$, $D_R \rightarrow D_R'$, $\Delta y \rightarrow \Delta y'$, $\Delta z \rightarrow \Delta z'$. Here the round-trip Doppler shift becomes 
\begin{align}
	\frac{\Delta \omega_{\textrm{e}}}{\omega_{\textrm{e}\mathcal{E}}} (t_{\mathcal{D}})= \, &  \epsilon_{\times}\frac{\Delta y \Delta z}{\Delta\ell^2} [ h_{\times}(ct_{{\mathcal{D}}} - x_{\mathcal{M}}) -\nonumber \\& \hspace{64pt} h_{\times}(ct_{{\mathcal{D}}} - 2 \Delta \ell- x_{\mathcal{S}})]  + \nonumber \\ & \epsilon_+ \frac{\Delta y^2 - \Delta z^2}{2\Delta\ell^2}  [ h_{+}(ct_{{\mathcal{D}}} - x_{\mathcal{M}})- \nonumber \\ &\hspace{64pt}h_{+}(ct_{{\mathcal{D}}} - 2 \Delta \ell- x_{\mathcal{S}})]\,. \label{eq:freq_shift}
\end{align}

Eq.~(\ref{eq:finalintensity}) is the main result of our work. It gives the instantaneous intensity measured at the end of the interferometry process in our toy model detector. By making $\bm{\epsilon} = 0$, we notice that the only non-vanishing term is the difference in cosines squared appearing in $I_{(\bm{0})}(t_{\mathcal{D}})$. These are the Minkowski contributions and, when $\widehat{\Theta}_{\mathcal{E}} = 0$, can be combined to give the usual single sine squared of the difference of the arms' lengths \cite{Maggiore2007}. 

The contributions of $T$ and $T'$ arise from the interaction of GWs with the laser beams and are all of linear order on the parameter $\bm{\epsilon}$. The first term in Eq.~($\ref{eq:non_trivial_pert}$) and the equivalent one for the other arm are the traditional perturbations obtained when discussing the detection of GWs and are associated with the phase difference of the interacting beams. In fact, they come from the phase of the initial conditions in Eqs.~(\ref{eq: x_comp_init_cond_ray_1}) and ($\ref{eq: x_comp_init_cond_ray_3}$) and are characterized mainly by the anisotropic change in the arms' radar lengths ($\Delta L$ and $\Delta L'$) which results in a difference of optical paths along them. We see that new effects are also present. One is proportional to the frequency shift arising from the time variation of the radar distance between the arm's extremities (cf. subsection \ref{subsec:radar_doppler_effect}) and the others are proportional to the initial value of the optical expansion parameter of the beams. Both were expected to influence such final intensity by their explicit presence on Eq.~(\ref{eq:intensity_evolution}), terms whose physical origins were previously brought to light (cf. subsection \ref{subsec:electriceq}). It is important to emphasize, as discussed in subsection \ref{subsec:freq_vs_radar_dist}, that the frequency shift does not contribute to any phase shift along the rays, since the phase on each ray is constant as the geometrical optics regime demands to be. Actually, this Doppler contribution simply originates from the electric field magnitude non-trivial propagation.

We note that although our experiment is set up in the dark fringe, the first curly bracket factor on Eq.~(\ref{eq:finalintensity}) informs us that if the unperturbed arms have equal lengths, \emph{i.e. $\Delta \ell = \Delta \ell'$}, then $I(t_{\mathcal{D}}) = 0$, even if GWs are present (and thus $\Delta L \neq \Delta L'$). In fact, in inertial frames of Minkowski spacetime, the interference pattern is quadratic in the difference of arms' lengths when it is small compared to the EM wavelength, and so a first conjecture is that, if this difference was only caused by the GW and yet the functional form of the intensity was the same $I(t_{\mathcal{D}}) \sim (\Delta L - \Delta L')^2$, there should be no contributions up to linear order in $\bm{\epsilon}$. Of course, as we derived, the form of the intensity changes itself by additional terms, but this property still holds in our toy model interferometer.

We seek now to compare the relevance of each of the contributions on Eq.~(\ref{eq:non_trivial_pert}). We will compare the terms on the $T$ function but a completely analogous analysis may be carried out for the $T'$ contributions. The amplitudes of the non-trivial contributions are: 
\begin{align}
	&C_1 \leqdef \frac{2\omega_{\textrm{e}\mathcal{E}}}{c} \Delta L(ct_{\mathcal{D}} - \Delta \ell), \\
	&C_2 \leqdef  \frac{\Delta \omega_{\textrm{e}}}{\omega_{\textrm{e}\mathcal{E}}}(t_{\mathcal{D}}), \\ &	
	C_{3a}\leqdef - \frac{2c \widehat{\Theta}_{\mathcal{E}}\Delta L(ct_{\mathcal{D}}- \Delta \ell) }{c\widehat{\Theta}_{\mathcal{E}} \Delta \ell + \omega_{\textrm{e}\mathcal{E}}}, 
	\\
	&C_{3b} \leqdef \frac{c X_P  \widehat{\Theta}_{\mathcal{E}} \Delta \ell  }{c\widehat{\Theta}_{\mathcal{E}} \Delta \ell + \omega_{\textrm{e}\mathcal{E}}} \epsilon_P h_P(ct_{\mathcal{D}} - 2\Delta \ell -x_{\mathcal{S}}),   
\end{align}
with $X_P$ symbolizing the coordinate factors:
\begin{align}
	X_+ \leqdef \frac{\Delta y^2 - \Delta z^2}{2\Delta \ell^2}\,, \quad X_{\times} \leqdef \frac{\Delta y \Delta z}{\Delta \ell^2}\,.
\end{align}

We begin by comparing  $C_1$ and $C_{3a}$:
\begin{align}
	\frac{|C_1|}{|C_{3a}|} &= \frac{\omega_{e \mathcal{E}}\Delta \ell}{c} \left( 1 + \frac{\omega_{\textrm{e}\mathcal{E}}}{c \widehat{\Theta}_{\mathcal{E}}\Delta \ell}\right) > \frac{\omega_{e \mathcal{E}}\Delta \ell}{c} = \frac{2 \pi \Delta \ell}{\lambda_{e \mathcal{E}}} \gg 1\,,
\end{align}
since we assume divergent beams, so that $\widehat{\Theta}_{\mathcal{E}}>0$. For aLIGO, $\mathcal{O}(C_1/C_{3a}) = 10^{11}$, remembering Eq.~(\ref{eq:init_Theta_est}) and that $\Delta \ell = 4$ km and $\omega_{e \mathcal{E}} = 1.8 \times 10^{15}$ rad/s \cite{Martynov2016}. Thus, $\mathcal{O}(C_1) \gg \mathcal{O}(C_{3a})$.

The remaining comparisons cannot be made in the same straightforward way.  Instead, we first assume our GW to be a general wave packet given by the Fourier decomposition:
\begin{equation}
	h_P(ct-x) = \textrm{Re}\bigg\{\int_{- \infty}^{\infty} \tilde{h}_P(\omega_g) \textrm{e}^{i\frac{\omega_g}{c}(ct - x)}d\omega_g \bigg\}, \label{eq:wave_packet}
\end{equation}
where $\textrm{Re}(\alpha)$ is the real part of $\alpha$. Then, by Eq.~(\ref{eq:freq_shift}), we know that a representative term of the frequency shift present in $C_2$ is of the form:
\begin{align}
	&\frac{\Delta \omega_{\textrm{e}}}{\omega_{\textrm{e}\mathcal{E}}}(t_{\mathcal{D}})  \sim \epsilon_P X_P [h_P(ct_{\mathcal{D}} -x_{\mathcal{S}}) - h_P(ct_{\mathcal{D}} - 2\Delta \ell -x_{\mathcal{S}})] \nonumber \\
	&= -2\epsilon_P X_P \int \sin\Big(\frac{\Delta \ell}{\lambdabar_g}\Big) \bigg[\textrm{Re}(\tilde{h}_P)\sin\bigg(\frac{ct_{\mathcal{D}} - \Delta \ell- x_{\mathcal{S}}}{\lambdabar_g}\bigg) \nonumber \\ & \hspace{75pt} +\textrm{Im}(\tilde{h}_P)\cos\bigg(\frac{ct_{\mathcal{D}} - \Delta \ell - x_{\mathcal{S}}}{\lambdabar_g}\bigg)\bigg]   d\omega_g\,,
	\label{eq:freq_modes}
\end{align}
where $\textrm{Im}(\alpha)$ is the imaginary part of $\alpha$. As for the radar distance perturbation in $C_1$, one may attain the contributions of each mode in the same fashion by writing:
\begin{align}
	&C_1   \sim
	\frac{\epsilon_P \omega_{\textrm{e}\mathcal{E}} X_P }{c} \int_{ct_{\mathcal{D}} -2\Delta \ell -x_{\mathcal{S}}}^{ct_{\mathcal{D}}-x_{\mathcal{S}}}\textrm{Re}\left\{\int \tilde{h}_P \textrm{e}^{i\frac{\omega_g}{c}w}d\omega_g \right\}dw
	\nonumber \\&  =2 \epsilon_P X_P \hspace{-3pt}\int \hspace{-1pt} \frac{\omega_{\textrm{e}\mathcal{E}}}{\omega_g} \sin\bigg(\frac{\Delta \ell}{\lambdabar_g}\bigg)  \bigg[\textrm{Re}(\tilde{h}_P) \cos\left(\hspace{-2pt}\frac{ct_{\mathcal{D}} - \Delta \ell - x_{\mathcal{S}}}{\lambdabar_g}\hspace{-2pt}\right)  \nonumber \\ & \hspace{75pt}-\textrm{Im}(\tilde{h}_P) \sin\left(\frac{ct_{\mathcal{D}} - \Delta \ell - x_{\mathcal{S}}}{\lambdabar_g}\right)  \bigg]d\omega_g\,. \label{eq:radar_dist_modes}
\end{align}

Factoring out the common parameters of $C_1$ and $C_2$, we may compare these two terms by looking at the integrands in Eqs.~(\ref{eq:freq_modes}) and (\ref{eq:radar_dist_modes}), which, apart from combinations of (bounded) harmonic functions, differ by a factor $\omega_{\textrm{e}\mathcal{E}}/\omega_g$. Since our description of the interferometry process is only valid in the electromagnetic geometrical optics regime and $\omega_g$ gives the scale of the metric variations (cf. Appendix \ref{app:geometrical_optics}), we must have, for each GW mode
\begin{align}
	\frac{|C_1|}{|C_2|} \sim \frac{\omega_{\textrm{e}\mathcal{E}}}{\omega_g} \gg 1\,.
\end{align}
In particular, for the aLIGO detectable spectrum, $\omega_e/\omega_g \geq 10^{11}$.

Contribution $C_{3b}$ may be rewritten as
\begin{align} &C_{3b} = \frac{cX_P\epsilon_P\widehat{\Theta}_{\mathcal{E}}\Delta \ell}{c\widehat{\Theta}_{\mathcal{E}}\Delta \ell + \omega_{\textrm{e}\mathcal{E}}}\textrm{Re}\left\{\int \tilde{h}_P \textrm{e}^{i\frac{\omega_g}{c}(ct_{\mathcal{D} - 2\Delta \ell - x_{\mathcal{S}}})}d\omega_g \right\} \nonumber \\ &= \frac{X_P\epsilon_P}{1 + \omega_{\textrm{e}\mathcal{E}}/c\widehat{\Theta}_{\mathcal{E}}\Delta \ell} \int \bigg[ \textrm{Re}(\tilde{h}_P) \cos \bigg(\frac{ct_{\mathcal{D}} - \Delta \ell - x_{\mathcal{S}}}{\lambdabar_g}\bigg)  \nonumber \\ &\hspace{35pt}-\textrm{Im}(\tilde{h}_P) \sin\left(\frac{ct_{\mathcal{D}} - \Delta \ell - x_{\mathcal{S}}}{\lambdabar_g}\right) \bigg]\cos\left(\frac{\Delta \ell}{\lambdabar_g}\right) \nonumber \\ 
& \hspace{13pt}+ \bigg[ \textrm{Re}(\tilde{h}_P) \sin \bigg(\frac{ct_{\mathcal{D}} - \Delta \ell - x_{\mathcal{S}}}{\lambdabar_g}\bigg) \nonumber \\ &\hspace{35pt}+ \textrm{Im}(\tilde{h}_P) \cos\left(\frac{ct_{\mathcal{D}} - \Delta \ell - x_{\mathcal{S}}}{\lambdabar_g}\right) \bigg]\sin\left(\frac{\Delta \ell}{\lambdabar_g}\right). \label{eq:Theta_modes}
\end{align}
Comparing this expression with Eq.~(\ref{eq:radar_dist_modes}), we conclude that the two last terms of Eq.~(\ref{eq:Theta_modes}) are always smaller than $C_1$, again because of the frequency ratio present in the latter and the fact that
\begin{equation}
	\frac{1}{1+\omega_{\textrm{e}\mathcal{E}}/c\widehat{\Theta}_{\mathcal{E}}\Delta \ell} < 1\,.
\end{equation}
As for the first two terms in $C_{3b}$, one arrives at the same conclusion, unless $\Delta \ell \approx n \lambda_g, n \in \mathbb{N}$,  because these modes are suppressed in $C_1$, but the corresponding ones in $C_{3b}$ are not.
Then, apart from these countable resonance GW frequencies that should not affect the overall integral, one concludes in the general case that
\begin{equation}
	\frac{|C_1|}{|C_{3b}|} \gg 1\,.
\end{equation}
In fact, on the particular case of the long-wavelength limit, $\Delta \ell /\lambda_g \ll 1$, we find:
\begin{equation}
\frac{\omega_{\textrm{e}\mathcal{E}}}{\omega_g}\sin{\left(\frac{\Delta \ell}{\lambdabar_g}\right)} \approx \frac{\omega_{\textrm{e}\mathcal{E}}}{\omega_g} \frac{\Delta \ell}{\lambdabar_g} = \frac{\Delta \ell}{\lambdabar_e} \gg 1\,,
\end{equation}
while
\begin{equation}
\frac{1}{1+\omega_{\textrm{e}\mathcal{E}}/c\widehat{\Theta}_{\mathcal{E}}\Delta \ell} \cos{\left(\frac{\Delta \ell}{\lambdabar_g}\right)} \approx 	\frac{1}{1+\omega_{\textrm{e}\mathcal{E}}/c\widehat{\Theta}_{\mathcal{E}}\Delta \ell} < 1\,.
\end{equation}

It is important to stress that for the above comparisons, we only assumed that the geometrical optics limit for light is valid and that the arms of the interferometer are much bigger than the laser wavelength. No assumption was necessary on the particular values of the parameters in question, allowing us to conclude, for any detector compatible with our modeling, that the dominant term is always the traditional one for normally incident GWs. We remember that our model assumes passive reflection and so it should be modified for the LISA detector. We also emphasize that the comparisons made were between the instantaneous values of each contribution and a more accurate treatment could be achieved by taking their time average.

\begin{figure}[t]
	\centering
	\includegraphics[scale=0.6]{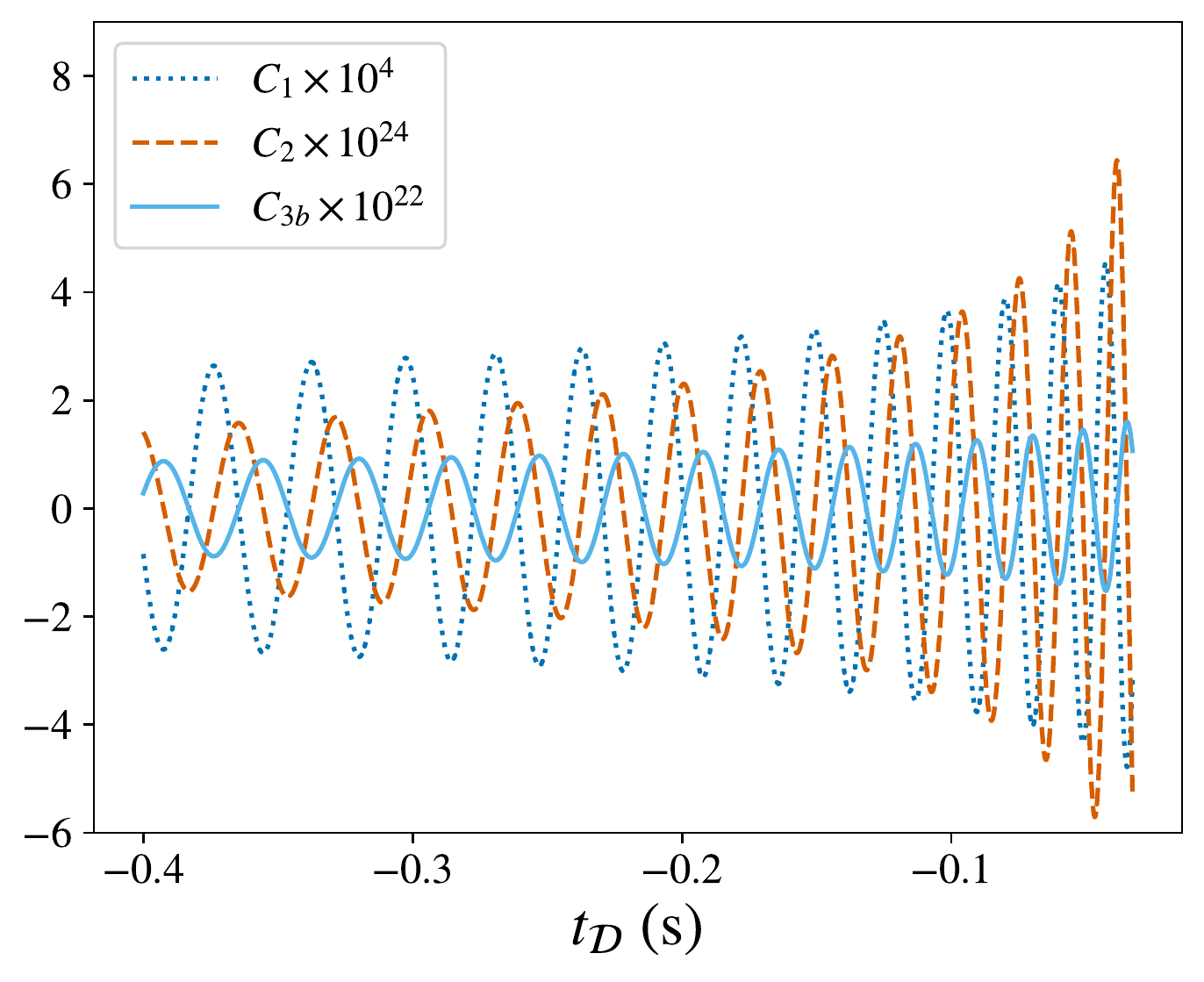}
	\caption{Contributions to the interference pattern as functions of the detection time for a typical GW signal observed by aLIGO corresponding to the inspiral phase of a binary source (up to $\omega_g \simeq 425$ rad/s). We have made $\theta = \pi/2$, $\phi = \pi/3$, $\Delta \ell =$ 4 km, $M_c = 28.3 \, \text{M}_{\odot}$, $r = 410$ Mpc and $\Phi_0=t_c=\iota = 0$.}
	 \label{fig:contributions}
\end{figure}

Although the above comparisons were made in a much more general case, we present on Fig.~\ref{fig:contributions} the amplitudes $C_1$, $C_2$ and $C_{3b}$ for the same simple template of the GW amplitude of Eq.~(\ref{eq:gw_binary}) and the same parameter values used in Fig.~\ref{fig:binary_waveform} basically corresponding to the aLIGO first detection. $C_{3a}$ was not shown since it is simply proportional to $C_1$. We note that, as anticipated, the non-traditional contributions can be safely neglected.

\section{Conclusion}
\label{sec:conclusion}

The main purpose of this work was to investigate two fundamental facets of the influence of GWs upon light, namely, the spatial and temporal perturbations induced on the null geodesics, and the modified electromagnetic field propagation along light rays. We assemble these aspects with particular attention, but not only, to how could they affect the interference pattern in a GW detector calculated from our newly derived Eq.~(\ref{eq:electric_evolution}) and estimate their quantitative relevance. The several contributions to this evolution equation are interpreted in terms of physical effects, that later will emerge in our final result. The whole approach relied on assuming light as a test field, satisfying the geometrical optics approximation of Maxwell's equations and on the linear regime of GWs around a flat background, not restricting ourselves to the GW long-wavelength limit. We described the idealized interferometer as part of the purely shearing TT reference frame through the covariant formalism of kinematic quantities, besides characterizing the light beams traveling along its arms, analogously, by their optical parameters.

We have computed the family of all possible null geodesics by integrating the constants of motion related with the symmetries of the GW spacetime, selecting from it those rays exchanged between any two TT observers, presented in Eqs.~(\ref{t_ray_1})--(\ref{z_ray_1}). This is achieved by the imposition of initial and partial boundary conditions, permitting the determination of the referred constants in terms of known experimental parameters, Eqs.~(\ref{A1}), (\ref{B1}), (\ref{delta}). While deriving such result, we were able to revisit and unite the central concepts of radar distance and frequency shift, respectively obtained in Eqs.~(\ref{radardistspheric}), (\ref{redshift}) and appearing in the later discussion of interferometry. Perturbations in light's spatial trajectories were shown not to disturb the radar distance between the observers, although the indiscriminate use of the hybrid model to simplify the description of light propagation was demonstrated to fail for predicting the behavior of other related quantities, such as the frequency along a ray. Although the radar distance and the round-trip frequency shift are non-infinitesimal quantities, used to describe interferometry beyond the long-wavelength limit, we clarified their relation with the infinitesimal shearing of the reference frame.

Finally, we considered both the curved nature of a non-monochromatic plane GW spacetime and the kinematics of the TT frame to evolve the electric field up to the detection event in an idealized Michelson-Morley-like interferometer, assuming passive reflection at the end mirrors. The result is found in Eqs.~(\ref{eq:Eray2}) and (\ref{M1})--(\ref{eq:M3}) for arbitrary GW incidence and detector configuration, being only necessary to impose further an initial condition that guarantees the tranversality of the electric field. Then, as our key result, we were able to compute, in Eq.~(\ref{eq:finalintensity}), for a normally incident GW, the final instantaneous EM interference pattern as a function of the GW amplitude, initial EM frequency, the laser beam divergence at emission, unperturbed arms' lengths and orientations. We found two new contributions besides the known traditional term related to the difference in optical paths: one associated to the frequency shift acquired by light during its round-trips in the arms, as a consequence of the TT kinematics, and other due to the expansion of the light beam. Despite being of linear order in the GW perturbative parameter $\bm{\epsilon}$, they showed to be negligible compared to the traditional contribution as long as the EM wavelength is much smaller than the GW one and the arms' lengths, conditions commonly understood as prerequisites for the validity of the geometrical optics approximation \cite{Schneider1992}. 
On the aLIGO case, we have estimated the value of the initial expansion parameter for the light beam and concluded that all non-traditional corrections are at most of order $10^{-11}$ when compared with the traditional one. A typical waveform for aLIGO was implemented to exemplify the behavior and magnitudes of the contributions.

A third new contribution to the interfered light intensity was foreseen to arise from the non-parallel transport of the EM polarization vector. For GW normal incidence we have shown that such vector is indeed parallel transported and, thus, this contribution is not present, but in a more general case, it is expected to perturb the measured signal. Moreover, ignoring such vector nature of the electric field in this case allows us to assess the interference pattern a priori, by only looking at the intensity evolution in the arms. For this, it is enough to consider, in view of Eq.~(\ref{eq: final_intensity_simplification}), that the electric fields before superposition at detection are given by
\begin{align}
E(t_{\mathcal{D}}) = \sqrt{I_{\mathcal{D}}}\,, \; E'(t_{\mathcal{D}}) = -\sqrt{I'_{\mathcal{D}}}\,,
\end{align}
and only use the intuitive Eq.~(\ref{eq:brightness_conservation}) to write the final intensity at one arm
\begin{align}
I_{\mathcal{D}} = I_{\mathcal{E}}\frac{\delta S_{\mathcal{E}}}{\delta S_{\mathcal{D}}}\left(\frac{ \omega_{\textrm{e}\mathcal{D}}}{\omega_{\textrm{e}\mathcal{E}}}\right)^2\,,
\end{align}
where its initial value is a function of the phase times a constant amplitude, i.e. $I_{\mathcal{E}} = \mathcal{E}^2\cos(\psi_{\mathcal{E}})$. So $I_T = [E(t_{\mathcal{D}}) + E'(t_{\mathcal{D}})]^2$ becomes, in this simplifying reasoning,
\begin{align}
I_T &= \mathcal{E}^2\Bigg[ \sqrt{\frac{\delta S_{\mathcal{E}}}{\delta S_{\mathcal{D}}}}\left(1 + \frac{\Delta \omega_{\textrm{e}}}{\omega_{\textrm{e}\mathcal{E}}}\right)\cos(\psi_{\mathcal{E}}) - \nonumber \\ & \hspace{65pt} \sqrt{\frac{\delta S_{\mathcal{E}}}{\delta S_{\mathcal{D}}'}}\left(1 + \frac{\Delta \omega_{\textrm{e}}'}{\omega_{\textrm{e}\mathcal{E}}}\right)\cos(\psi_{\mathcal{E}'})\Bigg]^2, \label{eq:intensity_conclusion}
\end{align} 
which in fact agrees with Eq.~(\ref{eq:finalintensity}), if one expands it up to linear order in $\bm{\epsilon}$ and relate, by Eq.~(\ref{Thetaandarea}), the ratio of areas with $\widehat{\Theta}$.

Another aspect studied is the origin of the frequency shift contribution to the final intensity pattern. By heuristic arguments, it is common to perceive that such a Doppler effect could play an independent role on the difference in phase of the interfered beams. We clarify why this is not the case, and from which aspect of the GW influence upon light a frequency related contribution could arise, as indeed occurs. In fact, Eq.~(\ref{eq:intensity_conclusion}) makes explicit that it comes from the evolution of the individual intensities along the arms.

Under the assumptions of this work, then, although several features regarding perturbations in light induced by GWs do ensue in linear order, e.g., spatial trajectory deviations, Doppler effect, polarization tilts and intensity fluctuations, one may certify that, at least for normal incidence, the detection of GWs justifiably relies, for quantitative purposes, on the interference pattern depending solely on the phase difference of the recombining rays, as presupposed throughout most of literature.

Further developments can be made in several directions. First, one could use the electric field here calculated to obtain the interference pattern for an arbitrary GW incidence, which we expect to have additional new contributions related with the polarization evolution. Second, it is possible to assume the interferometer to be in a reference frame other than the TT one with, in general, different kinematic features. A third viable option is to discuss the themes here studied beyond the geometrical optics regime for light, but still in the context of interferometry.

\acknowledgments
J. C. L. thanks Brazilian funding agency CAPES for PhD scholarship 88887.492685/2020-00. I. S.M. thanks Brazilian funding agency CNPq for PhD scholarship GD 140324/2018-6.

\appendix

\section{On the domains of the $\epsilon$-parametrized light rays}
\label{app:domains}

Each one of the outgoing null geodesic arcs from Eq.~(\ref{consistent_models}) is a function
\begin{align}
\xi_{(\bm{\bm{\epsilon}})}: & \, D_{(\bm{\bm{\epsilon}})} \rightarrow \mathbb{R}^4\,,
\end{align}
whose domain is
\begin{align}
D_{(\bm{\epsilon})} \leqdef [0, \vartheta_{\mathcal{R}}]\,, \label{domains}
\end{align}
and which satisfies the discussed mixed conditions related to Eq.~(\ref{mixed_conditions}). More precisely, thinking about the parametrized curves as functions of $(\bm{\bm{\epsilon}}, \vartheta, \bm{P})$, one can evaluate $\xi_{(\bm{\bm{\epsilon}})}^i$ at its final event, substituting the spatial coordinates of $\mathcal{R}$, $x^i_{\mathcal{M}} \in \bm{P}$, and then solving for the value $\vartheta_{\mathcal{R}}$:
\begin{align}
\xi_{(\bm{\bm{\epsilon}})}^i(\vartheta_{\mathcal{R}}) = x^i_{\mathcal{M}} \; \Rightarrow \; \vartheta_{\mathcal{R}} = f(\bm{\bm{\epsilon}}, \bm{P})\,. \label{vartheta}
\end{align}
So $\mathcal{R}$ and, consequently, $\vartheta_{\mathcal{R}}$ are determined by both the parameters $\bm{P}$ and $\bm{\bm{\epsilon}}$. This is why $D_{(\bm{\bm{\epsilon}})}$ depends on $\bm{\bm{\epsilon}}$. Of course, since Eq.~(\ref{domains}) holds for all $\bm{\epsilon}$ including $\bm{0}$, the first term of the expansion, $f(\bm{0}, \bm{P})$, will be equal to what we call $\vartheta_{\mathcal{R}_{(0)}}$ in the main text.

Also, for all $\vartheta \in D_{(\bm{\epsilon})} \cap D_{(\bm{0})}$, it is true that
\begin{align}
\xi_{(\bm{\epsilon})}(\vartheta) = \xi_{(\bm{0})}(\vartheta) + \epsilon_P\,\xi^P(\vartheta)\,, \label{param_curve}
\end{align}
from which one derives Eq.~(\ref{k_expansion}). The above expansion consists on splitting the functional dependence of the curve into that of the model $\mathbb{M}_{(\bm{0})}$ plus some additional terms.

Of course we can still write Eq.~(\ref{param_curve}) for all points in $D_{(\bm{\epsilon})}$ as long as we extend the unperturbed geodesic arc to $\xi_{(\bm{0})|\textrm{ext}}$, maintaining its geodesic character, to this domain (or even to $\cup_{\bm{\epsilon}} D_{(\bm{\epsilon})}$). However, in this case, one should be aware that, while evaluating Eq.~(\ref{param_curve}) in $\vartheta_{\mathcal{R}}$, the first term, although written with a subscript $(\bm{0})$, will have contributions depending on $\bm{\epsilon}$, since, from Eq.~(\ref{vartheta}):
\begin{align}
\xi_{(\bm{0})|_{\textrm{ext}}}(\vartheta_{\mathcal{R}}) &= \xi_{(\bm{0})}(\vartheta_{\mathcal{R}_{(\bm{0})}}) + k_{(\bm{0})}(\vartheta_{\mathcal{R}_{(\bm{0})}})\frac{\partial f}{\partial \epsilon_P}(\bm{0}, \bm{P}) \epsilon_P\,. 
\end{align}

\section{Christoffel symbols}
\label{app:christoffel_symbols}
The Christoffel symbols of the metric (\ref{metric}) can be simplified, when dealing with approximations up to linear order in $\bm{\epsilon}$, to:
\begin{align}
	\Gamma^{t}_{\beta \gamma} &= \frac{\epsilon_P}{2} h^P_{\beta \gamma,t}\,, \label{gamma^t}\\
	\Gamma^{i}_{t j} &= \frac{\epsilon_P}{2} h^P_{j i,t}\,, \label{gamma_t}\\
	\Gamma^{i}_{x j} &= \frac{\epsilon_P}{2} h^P_{j i,x}\,, \label{gamma_x}\\
	\Gamma^{i}_{y j} &= \epsilon_P h^P_{y [i,j]}\,, \label{gamma_y}\\
	\Gamma^{i}_{z j} &= \epsilon_P h^P_{z [i,j]}. \label{gamma_z}
\end{align}
\section{Geometrical optics approximation of Maxwell equations}
\label{app:geometrical_optics}

The geometrical optics approximation of Maxwell's equations in vacuum,
\begin{subequations}
	\label{eq:maxwell}
	\begin{eqnarray}
		{F^{\mu\nu}}_{;\nu} &=& 0\,, \label{eq:gauss} \\
		F_{[\mu\nu;\lambda]} &=& 0\,,  \label{eq:faraday}
	\end{eqnarray}
\end{subequations}
is established by searching for solutions of these field equations in the form of a one-parameter ($\eta$) 
family of electromagnetic fields \cite{Ehlers1967, Misner1973, Schneider1992, Perlick2000, Ellis2012, Harte2019}:
\begin{subequations}
	\label{eq:ansatz}
	\begin{eqnarray}
		F_{\mu \nu}(x, \eta) &=& f_{\mu \nu}(x, \eta) \textrm{e}^{i \psi(x)/\eta}\,, \label{eq:ansatz_product} \\
		f_{\mu \nu}(x, \eta) &\leqdef& \sum_{n = 0}^N f_{(n)\mu \nu}(x) \eta^n \quad (N \ge 0)\,. 	\label{eq:amplitude}
	\end{eqnarray}
\end{subequations}

\begin{table*}[t]
	\centering
	\begin{tabular}{ccc}
		Order & Final equation from (\ref{eq:gauss}) & Final equation from (\ref{eq:faraday})\\
		\hline \\
		$\eta^{-1}:$ & ${f_{(0)}}^{\mu\nu}\tilde{k}_\nu = 0$ &  $f_{(0)[\mu\nu}\tilde{k}_{\lambda]}=0$ \\
		\\
		\hline \\
		$\eta^p\quad (0\leq p\leq N-1):$ &  ${{f_{(p)}}^{\mu\nu}}_{;\nu} + i{f_{(p+1)}}^{\mu\nu}\tilde{k}_\nu = 0$   & $f_{(p)[\mu\nu;\lambda]} + if_{(p+1)[\mu\nu}\tilde{k}_{\lambda]} = 0$ \\
		\\
		\hline \\
		$\eta^N$: &  ${{f_{(N)}}^{\mu\nu}}_{;\nu} = 0$   & $f_{(N)[\mu\nu;\lambda]} = 0$
	\end{tabular}
	\caption{\label{tab:hierarchy}Hierarchy of Maxwell's equations for the geometrical optics approximation.}		
\end{table*}

{In general, $f_{\mu\nu}$ is a complex antisymmetric smooth tensor field 
	and $\psi(x)/\eta$ is a real smooth scalar field; these are called, respectively, the amplitude 
	and phase of the electromagnetic wave; $\eta$ is a dimensionless perturbation parameter proportional to the 
	wavelength of the electromagnetic wave. Naturally, the real part of $F_{\mu \nu}$ must be taken in the end. This \emph{Ansatz} 
	generalizes the plane wave monochromatic solution of Maxwell's equations in Minkowski spacetime (in pseudo-Cartesian 
	coordinates, adapted to  an inertial frame of reference), and is expected to represent, in the limit $\eta 
	\to 0$, a rapidly oscillating function of its phase, with a slowly varying amplitude. Moreover, the vector field 
	defined by
	\begin{equation} 
		\label{wave-vector}
		\tilde{k}_{\mu}(x) \leqdef  \psi_{,\mu}(x)
	\end{equation} 
	is supposed to have no zeros in the considered region (irrespective of the values of $\eta$), and 
	\begin{equation}
		\label{physical_wavevector}
		k_\mu \leqdef \frac{\tilde{k}_\mu}{\eta}
	\end{equation}
	should be interpreted as the wave vector field of the electromagnetic wave, proportional to the momentum field of a 
	stream of photons. Finally, $f^{(0)}_{\mu \nu}$ is assumed to vanish  at most in a set of measure zero. 
	Inserting Eq.~(\ref{eq:ansatz}) into Maxwell's equations (\ref{eq:maxwell}) and demanding their validity for all 
	values of $\eta$, we find $N + 1$ hierarchical relations, the first two of them given by (cf. Table 
	\ref{tab:hierarchy}):
	\begin{itemize}
		\item dominant $\eta^{-1}$ order:
	\end{itemize}
	\begin{subequations}
		\label{eta-1}
		\begin{eqnarray}
			{f_{(0)}}^{\mu \nu} \tilde{k}_\nu   & = & 0\,,\label{eq:gauss-1} \\
			f_{(0)[\mu \nu} \tilde{k}_{\lambda]} & = & 0\,, \label{eq:faraday-1}
		\end{eqnarray}
	\end{subequations}
	and
	\begin{itemize}
		\item subdominant $\eta^0$ order:
	\end{itemize}
	\begin{subequations}
		\label{eq:epsilon0}
		\begin{eqnarray}
			{{f_{(0)}}^{ \mu \nu}}_{;\nu} + i \tilde{k}_\nu {f_{(1)}}^{\mu \nu} & = & 0\,, \label{eq:gauss0} \\
			f_{(0)[\mu \nu;\lambda]} + i  f_{(1)[\mu \nu}\tilde{k}_{\lambda]} & = & 0\,. \label{eq:faraday0}
		\end{eqnarray}
	\end{subequations}
	
	Projecting Eq.~(\ref{eq:faraday-1}) onto $\tilde{k}^{\mu}$ and taking Eq.~(\ref{eq:gauss-1}) into account, it 
	immediately follows that
	\begin{equation}
		\label{eq:null_tangent}
		\tilde{k}_{\mu} \tilde{k}^{\mu} = 0 = k_\mu k^\mu\,,
	\end{equation}
	which implies that the integral curves of the wave vector field (or rays) are null curves, and, 
	together with Eq.~(\ref{wave-vector}), that the surfaces of constant phase are null hypersurfaces. 
	Besides, since $\tilde{k}^{\mu}$ has vanishing curl, these equations also show that the rays are geodesics, 
	also demonstrating that the light rays form a bundle with zero optical vorticity. 
	These are consistent with the results one obtains when studying the characteristic surfaces and 
	bi-characteristic curves of Maxwell's equations in vacuum \cite{Lichnerowicz1960}, or even when considering 
	shock waves of the electromagnetic field \cite{Papapetrou1977}.
	
	Besides, projecting Eq.~(\ref{eq:faraday0}) onto $\tilde{k}^{\mu}$, and using Eqs.~(\ref{eq:gauss-1}), (\ref{eq:faraday-1}), 
	(\ref{eq:gauss0}) and (\ref{eq:null_tangent}), we get
	\begin{equation}
		\label{eq:transport_zero_order}
		f_{(0)\mu \nu;\lambda}\tilde{k}^{\lambda} + \frac{1}{2} {\tilde{k}^{\lambda}}_{;\lambda}  f_{(0)\mu \nu} = 0\,.
	\end{equation}
	Here, we note that the above equation gives the evolution of $f^{(0)}_{\mu \nu}$ independently of 
	the other $f^{(p)}_{\mu \nu}$ ($p = 1, ..., N$). The usual geometrical optics approximation relies on taking $ \eta \to 0 $, and assuming 
	that $f^{(0)}_{\mu \nu}$ is a good approximation for the amplitude of the electromagnetic field, in which case all other
	contributions may be disregarded. This assumption 
	translates the physical demand that $F_{\mu\nu}(x,\eta)$ is to vanish at an arbitrarily large 
	discrete number of hypersurfaces (``nodes''), so that it can be interpreted as a realistic wave.
	If we stick to it, $F_{\mu \nu} (x, \eta) \approx f_{(0)\mu \nu}(x) \textrm{e}^{i \psi(x)/\eta}$, and the
	higher-order corrections are to be neglected \cite{Harte2019}}. Then, Eqs.~(\ref{eta-1}) and (\ref{eq:transport_zero_order}) become, 
respectively, equivalent to
\begin{subequations}
	\begin{eqnarray}
		k_\nu F^{\mu \nu} & = 0\,, \label{faraday-eigen} \\
		k_{[\lambda} F_{\mu \nu]} & = 0\,, \label{hodge-eigen}
	\end{eqnarray}
\end{subequations}
and
\begin{equation}
	\label{faraday-transport-equation}
	F_{\mu \nu;\lambda}k^{\lambda} + \frac{1}{2} {k^{\lambda}}_{;\lambda} F_{\mu \nu} = 0\,,
\end{equation}
where we have already included $\eta$ in the previous two equations, since they both contain the same orders of $\tilde{k}_\mu$.
Eqs.~(\ref{faraday-eigen}) and (\ref{hodge-eigen}) show that the wave vector is a principal 
null direction of both the electromagnetic field and its dual, and the considered \emph{Ansatz} 
corresponds (approximately) to a null electromagnetic field \cite{Hall2004, Harte2019}. 
Last, Eq.~(\ref{faraday-transport-equation}) shows how the electromagnetic field is transported 
along any of its associated light rays, and, together with Eq.~(\ref{hodge-eigen}), is the path 
leading to the transport equation for the electric field appearing in \cite{Santana2020} and to the results presented here therefrom.

Using the above constraints, we can establish a condition for the validity of the geometrical optics regime in the particular spacetime we use throughout this series, namely, a GW perturbed Minkowski background. For this, we replace $F_{\mu \nu} (x, \eta) \approx f_{(0)\mu \nu}(x) \textrm{e}^{i \psi(x)/\eta}$ in Eq.~(\ref{eq:gauss}) and find:
\begin{equation}
	\frac{i\tilde{k}_{\nu}f_{(0)}^{\mu \nu}}{\eta} + f_{(0),\nu}^{\mu \nu}  + \Gamma^{\nu}_{\alpha \nu} f_{(0)}^{\mu \alpha}= 0\,. \label{eq:go_app}
\end{equation}
In the particular case of our GW spacetime, in addition to $1/\eta$, there are two other expansion parameters $(\epsilon_+, \epsilon_{\times})$ that need to be taken into account simultaneously. In order to obtain the hierarchical relations (\ref{eq:gauss-1}), we need to check whether the last two terms in the above equation can be neglected as compared to the first one. Indeed, one of the geometrical optics assumptions is that the amplitude of the Faraday tensor varies much less than its phase and thus the second term (both imaginary and real parts) is considered to be much smaller than the first one. Furthermore, remembering that
\begin{align}
	\tilde{k}_{\nu} = \tilde{k}_{\nu (\bm{0})} + \epsilon_P\tilde{k}^P_{\nu} = \eta k_{\nu}\,,
\end{align}
Eq.~(\ref{eq:go_app}) can be written as having contributions of three orders, namely
\begin{align}
	\left[\left(\frac{1}{\eta}\right)i\tilde{k}_{\nu(\bm{0})} + \left(\frac{\epsilon_P}{\eta}\right)i\tilde{k}_{\nu}^P + \Gamma^{\alpha}_{\nu \alpha}\right]f_{(0)}^{\mu \nu} \approx 0\,, \label{eq:geo_opt_app}
\end{align}
where the Christoffel symbols are of order $\epsilon_P\omega_g$ for each mode in Eq.~(\ref{eq:wave_packet}). Since $\mathcal{O}(\tilde{k}^P/\eta) = \omega_{\textrm{e}\mathcal{E}}$, the crossed term is of order $\epsilon_P\omega_{\textrm{e}\mathcal{E}}$ and, therefore, we must have
\begin{align}
	\frac{\omega_{\textrm{e}\mathcal{E}}}{\omega_g} \gg 1\,, \label{eq:freq_ratio}
\end{align}
so that the last term is negligible compared to the others and thus the transversality of the EM field under the geometrical optics regime and all of its consequences are guaranteed. Of course a more rigorous and similar argument can be made by comparing the order of magnitude of the contributions in the real and imaginary parts of Eq.~(\ref{eq:geo_opt_app}), which would ultimately lead to the same conclusions. Condition (\ref{eq:freq_ratio}) is in agreement with the usual statement that the validity of the geometrical optics limit resides in EM wavelengths much smaller than the other relevant lengths of the system in question.


%

%

\end{document}